\documentclass[twocolumn,showpacs,pre]{revtex4} %\documentclass[preprint,showpacs,pre]{revtex4}
\usepackage{epsfig,graphicx,psfrag,color}\usepackage{dcolumn}\usepackage{amsmath,amssymb}\usepackage{deleq}
\begin{document}

\title{Dynamics of grain ejection by sphere impact on a granular bed}
\author{S. Deboeuf$^{1,2}$}
\author{P. Gondret$^1$}
\author{M. Rabaud$^1$}
\affiliation{$^1$Univ. Paris-Sud, Univ. Paris 6, CNRS, Lab. FAST, UMR 7608, B\^at. 502, Campus Univ, 91405 Orsay, France}
\affiliation{$^2$Univ. Paris 6, Univ. Paris 7, CNRS, Lab. de Physique Statistique de l'Ecole Normale Sup\'erieure, UMR 8550, 24 rue Lhomond, 75231 Paris Cedex 05, France}

\date{\today}

%\pacs{45.70.-n}{Granular systems} 
%\pacs{45.50.-j}{Dynamics and kinematics of a particle and system of particles}
%\pacs{83.80.Fg}{Granular solids}
%\pacs{96.15.Qr}{Impact phenomena}

\begin{abstract}
The dynamics of grain ejection consecutive to a sphere impacting  a granular material is investigated experimentally and the variations of the characteristics of grain ejection with the control parameters are quantitatively studied. The time evolution of the corona formed by the ejected grains is reported, mainly in terms of its diameter and height, and favourably compared with a simple ballistic model. A key characteristic of the granular corona is that the angle formed by its edge with the horizontal granular surface remains constant during the ejection process, which again can be reproduced by the ballistic model.  The number and the kinetic energy of the ejected grains is evaluated and allows for the calculation of an effective restitution coefficient characterizing the complex collision process between the impacting sphere and the fine granular target. The effective restitution coefficient is found to be constant when varying the control parameters.   
\end{abstract}

\maketitle
\parindent 0pt

\section{Introduction} \label{sec_intro}
Impact cratering has been recognized as an important geologic process for the last decades when the lunar craters have been finally attributed to impact structures rather than giant volcanoes as believed until 1950's~\cite{melosh}. 
 The planetary impact craters such as the ones observed commonly on the Moon or the Earth result from very high energy impacts of meteorites and thus involve numerous and very complex phenomena such as shock and rarefaction wave propagation, melt and vaporization of the projectile and target materials, together with excavation by displacement and ejection of the target material~\cite{melosh}.  
 In light of evidences of the discrete and sandy nature of planet surface, laboratory scale experiments of high energy impacts on granular materials were conducted~\cite{cook, yamamoto}.
 Since a few years, physicists have conducted laboratory scale experiments with rather low impact energy on granular matter, interesting in the crater morphology and searching for scaling laws for the crater size~\cite{debruyn03, durian03, zheng, debruyn07}. Even though their energies are typically many orders of magnitude smaller than those of meteorite impacts, these small scale experiments on granular impacts may be relevant to planetary impact processes,  as the progression of crater morphologies as a function of impact energy has been shown to mirror that seen in lunar craters~\cite{debruyn03}. 
  In these impact experiments, physicists have also been interested in the penetration of the impacting sphere in the granular target~\cite{debruyn04, ciamarra, durian05, hou, durian07, goldman, seguin}. Indeed, despite recent progress on the complex rheology of granular matter~\cite{GDRMidi}, the penetration dynamics of a solid sphere into a granular medium is still difficult to understand well as it involves both the complex drag resulting from frictional and collisional processes, and the final stop involving the complex ``liquid/solid" transition exhibited by granular matter~\cite{mills08}. The penetration dynamics of the impacting sphere and the grain ejection have been shown to be very different when the granular material is not dense but loose: A spectacular thin granular jet can raise very high after the impact as first demonstrated in Refs~\cite{cook, thoroddsen, lohse}. The effects on this granular jet of the interstitial fluid~\cite{royer, caballero} and of the initial packing fraction of the target~\cite{marston}  have then been studied. In the dense case, no granular jet but a growing granular corona is seen after the impact. These different kinds of grain ejection can be related to similar kinds of liquid ejection consecutive to the impact of a droplet into a deep or thin layer of liquid that have been first filmed by~\cite{worthington} and then studied extensively~\cite{fedorchenko, yarin}. Much less studies focusing on the grain ejection have been performed in the dense granular case: Ogale {\it et al.}~\cite{ogale} have measured the mass of the spilled-over grains and Boudet {\it et al.}~\cite{boudet} have proposed a model of ejection from the crater growth in a layer of thickness small compared to the size of both projectile and target grains at low impact velocities ($\sim$ 1~m/s), while Cintala {\it et al.}~\cite{cintala} measured  ejection speeds and angles of grains at high impact velocities ($\sim$ 1 km/s). 
Another type of  experiments concern the impact of one bead with a granular target made of the same beads~\cite{beladjine}.     
  Such impacts and grain ejections by an impacting projectile have been recently simulated in limiting cases, for which the impact energy is very low or very high~\cite{tsimring, wada} or for which the projectile size is about the grain size~\cite{oger, crassous, bourrier}. 
We focus in the present paper on the dynamics of the granular corona formed by the ejected grains from a dense and deep pile upon low speed ($\sim$ 1~m/s) impacts. In section~\ref{sec_exp}, the experimental setup is described together with the measurements. A simple ballistic model is then presented in section~\ref{sec_mod} and compared to the experimental data. The results are discussed in section~\ref{sec_disc} and section~\ref{sec_concl} ends this paper with a conclusion.

\section{Experiments} \label{sec_exp}
Each experimental run consists of dropping a solid sphere into a granular medium.
Four different steel spheres of density $\rho_s$=$7800$~kg/m$^3$  are used as impactors, with different radius $R = 5.15, 6.75, 7.55, 9.50$~mm and masses $M$ ranging thus from $4.5$~g to $30$~g.
The steel sphere is initially hold by a magnet through a semi-spherical hole, so that the sphere can be dropped without any translational nor rotational velocity by pulling up the magnet.  The sphere is released directly above the center of a container and falls along its axis. Being dropped from the height $h$ above the granular surface which is varied between $8$~cm and $60$~cm,  the sphere thus impacts the granular material with the velocity $U_c$ varying from $1$~m/s to $4$~m/s and energy $E_c=Mgh$  ranging from $3.10^{-3}$~J to $2.10^{-1}$~J.
The target material consists in sieved glass beads of density $\rho_g=2500$ kg/m$^3$ ($\rho_s/\rho_g=3.1$) and mean diameter $2r=0.4\pm0.1$~mm ($R/r\sim25-50$), thus of mass $m=8.10^{-5}$~g ($M/m\sim10^4-10^5$). 
 The granular material fills the cylindrical container of diameter $19$ cm and height $26$ cm. The size ratio of the container diameter over the sphere diameter is always larger than $10$, so that there is no influence of the radial confinement by the lateral walls of the container neither of the bottom wall~\cite{seguin} as the height of the packing is large. Before each drop, the granular medium is prepared by gently stirring the grains with a thin rod. The container is then overfilled and the surface levelled using a straightedge. The typical value of the solid volume fraction of the packing is $60 \%$. Each  impact experiment is lighted from the front to enhance contrast between the grains and the black background, and side view images are recorded by a high speed camera at  the rate of $500$~images/s and resolution of $0.16$~mm/pixel. The $256$ gray level images are thresholded to identify grains.

%% Figure 1
\begin{figure}
\centerline{ \includegraphics[width=.75\linewidth]{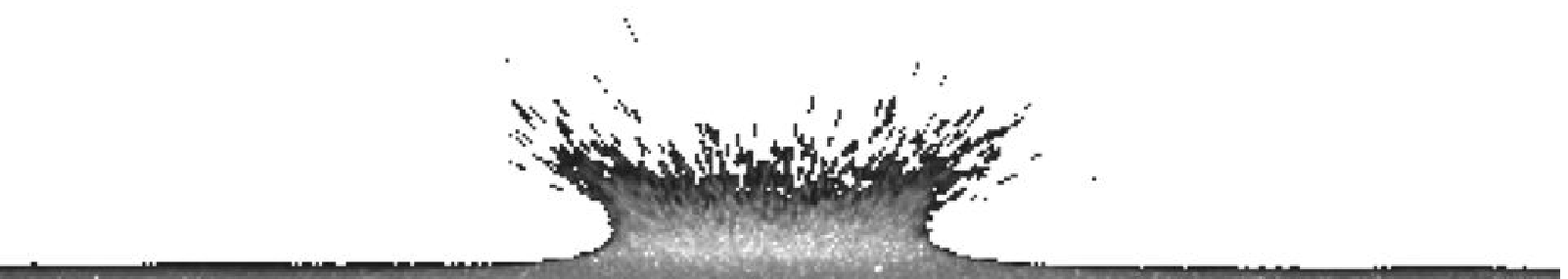} }
\centerline{ \includegraphics[width=.75\linewidth]{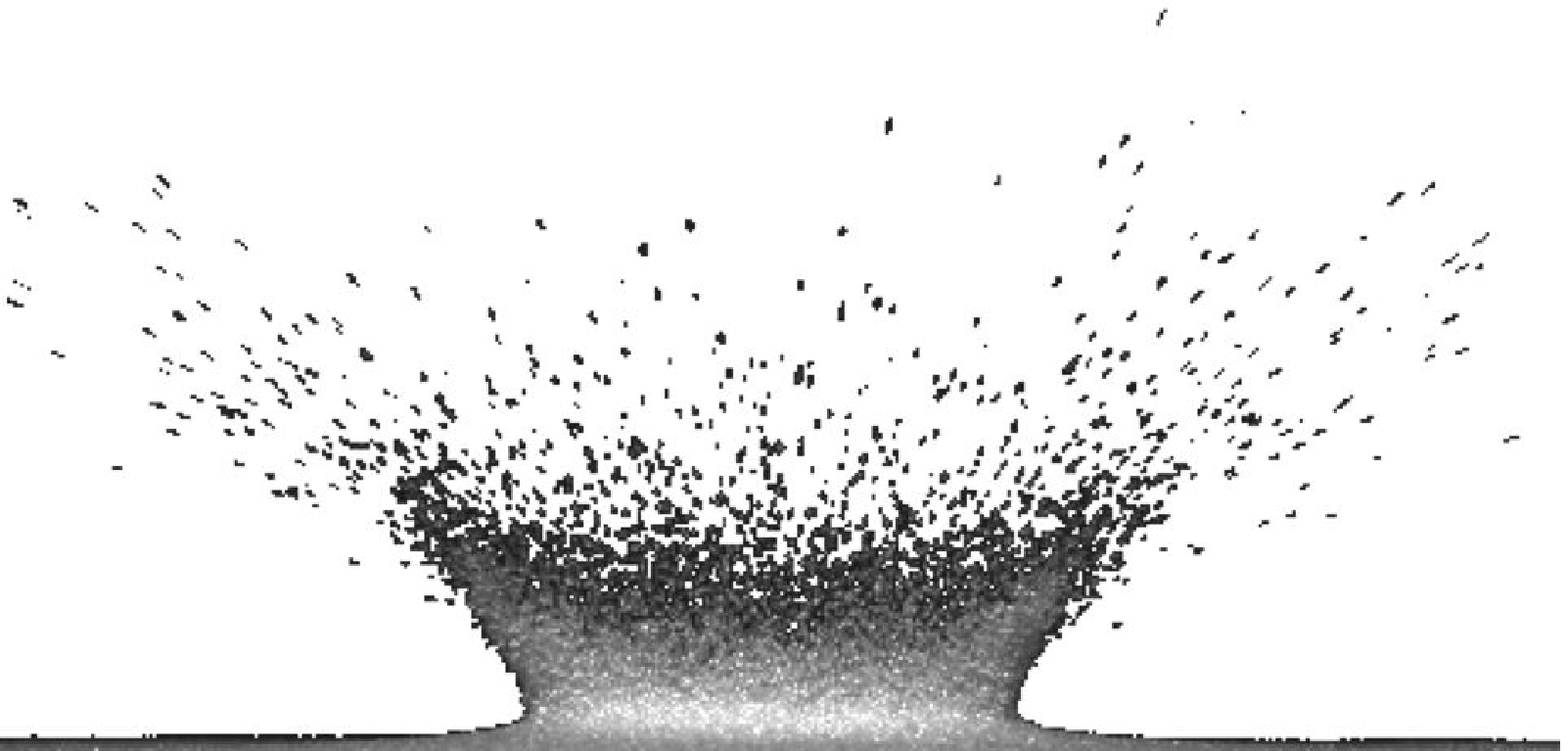} }
\centerline{ \includegraphics[width=.75\linewidth]{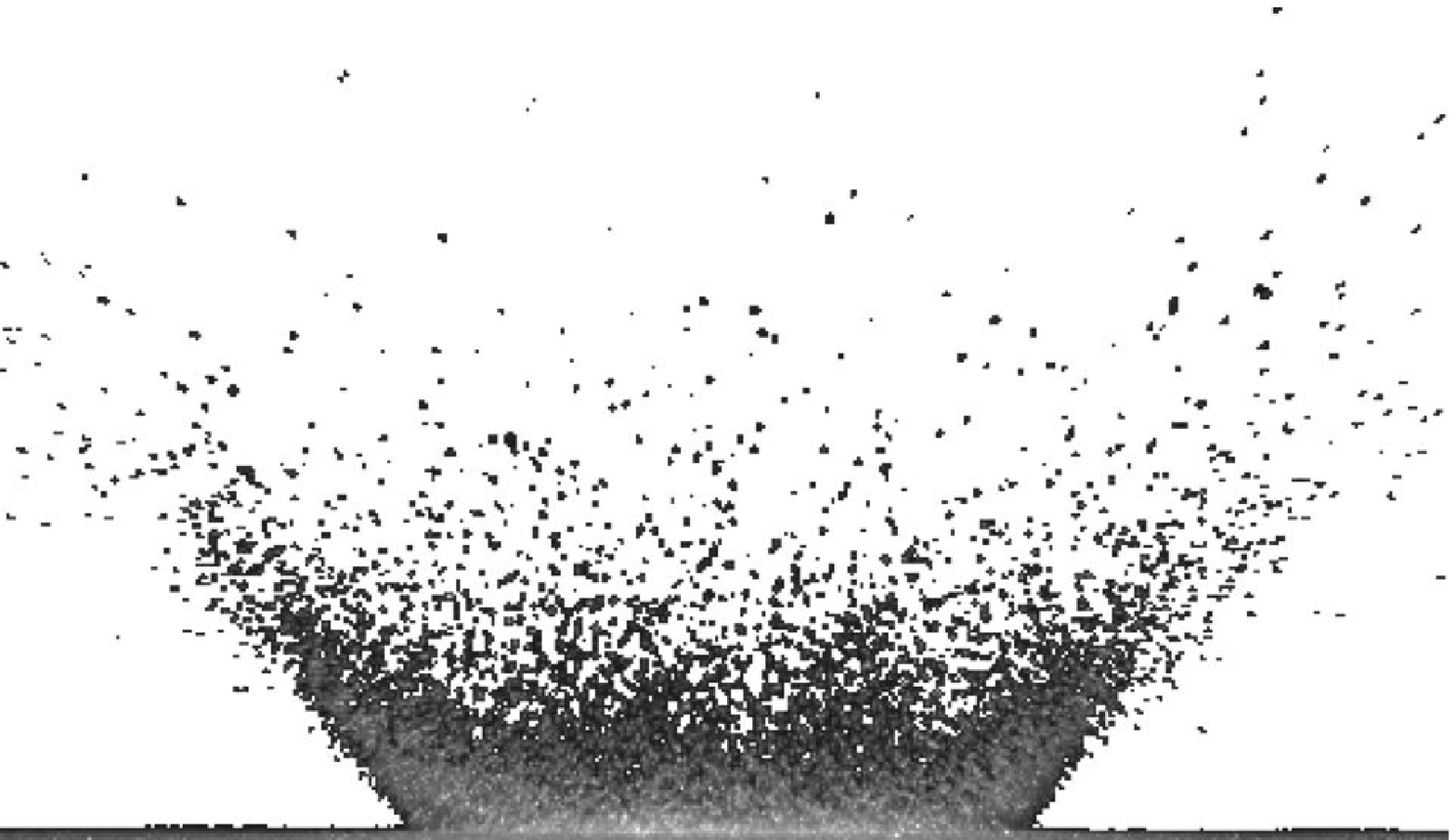} }
\centerline{ \includegraphics[width=.75\linewidth]{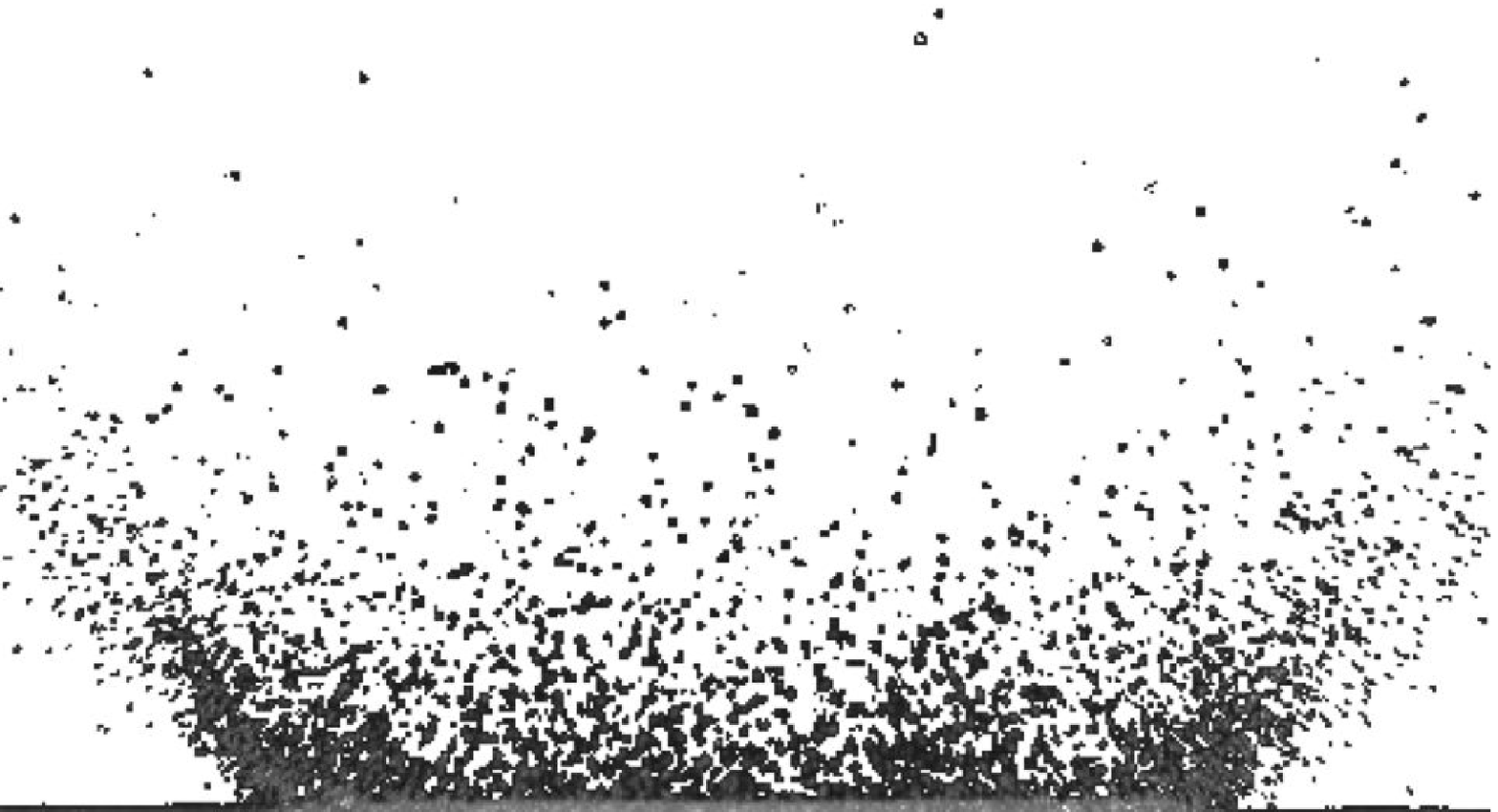} }
\caption{Sequence of side view images separated by $30$~ms, showing the typical time evolution of ejected grains after the impact of a steel sphere of radius $R = 6.75$ mm dropped from the height $h=30$ cm ($U_c=2.45$~m/s, $E_c=30$~mJ). }
\label{Figseq}
\end{figure}

\begin{figure}
\centerline{ \includegraphics[width=.75\linewidth]{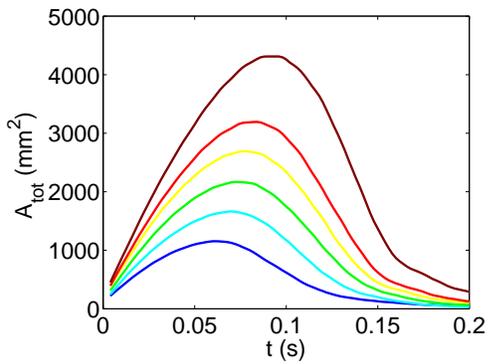} }
\caption{(Color online) Time evolution of the total apparent surface of ejected grains $A_{tot}$ for an impacting steel sphere of radius $R = 7.55$ mm dropped from different heights $h=13$, $23$, $33$, $43$, $48$ and $58$~cm (from bottom to top).}  
\label{Figatot}
\end{figure}

Figure~\ref{Figseq} shows a sequence of side view images separated by $30$ ms illustrating the ejection dynamics of the grains after the sphere impact  defined as the time $t = 0$. One can see few isolated ejected grains above a dense corona of grains that expands radially and vertically.
The amount of ejected grains can be quantified by measuring the apparent surface area $A_{tot}$ of all the grains in each image. The evolution of $A_{tot}$ as a function of time is reported in Fig.~\ref{Figatot} for several experiments corresponding to different dropping heights $h$ of the same sphere. After the impact at $t = 0$ defined as the first contact between the bottom of the sphere and the granular surface,  $A_{tot}$ increases up to a maximal value denoted $A_{tot\,max}$ at the time $t_{A_{tot\,max}}$ before decreasing.  Each curve corresponds to a single impact experiment, thus without any ensemble averaging, but with a little smoothing by a slide-average over a time window of $10$~ms so that no data appears before $t = 4$~ms in the reported figures. Increasing the dropping height $h$ of a given sphere, the values $A_{tot\,max}$ and $t_{A_{tot\,max}}$ increase, accounting for the  increase of both the number of ejected grains and the dynamics duration. The same goes when keeping constant the dropping height $h$ and increasing the mass $M$ of the impacting sphere.
It is worthnoting that the exact relation between the measured apparent surface area $A_{tot}$ of the ejected grains and the real total number of ejected grains $N_{ej}$ is not straightforward. For dilute zones of ejected grains, all the grains can be seen in the 2D images, but there is still the problem of focusing: A far grain appears smaller than a close grain so that the grains do not have the same apparent area. The most problematic case concerns however the dense zones where some grains can be hidden by other grains. 
Using the fact that the ejection process is axisymmetric, the number of ejected grains $N_{ej}$ is related to the measured apparent surface area $A_{tot}$ by: 
\begin{eqnarray}
N_{ej} = \overline{n} w \pi A_{tot\,max}  \mathrm{ , } \label{eq_nej} 
\end{eqnarray}
where $w$ is the unknown corona thickness and $\overline{n}$ the unknown  number of grains per unit volume. 
The unknown product $\overline{n} w$ will be deduced in the following by comparing our scaling laws for the grain ejection with scaling laws already known in the literature for the crater size.

%% Figure 2
%%

%% Figure 3
\begin{figure}
\centerline{ \includegraphics[width=\linewidth]{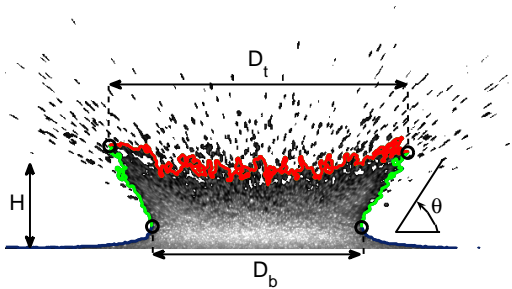} }
\caption{(Color online) Geometrical parameters characterizing the corona of ejected grains.}
\label{Figsetup}
\end{figure}

In impact experiments of a drop onto a liquid layer, a beautiful corona is classically observed~\cite{worthington, yarin}. In such a liquid case, the corona is easy to define and extract as the liquid is a continuous medium. In the granular case, the corona is less easy to define as the ejected grains are individual entities. The granular corona is here defined as the largest connected part of grains in each side view image. We have checked that the size of the corona does not depend significantly on the lighting and contrast of the images. To investigate the time evolution of the corona and characterize its shape which is basically axisymmetrical, its contour is extracted for each image. An example of such a contour is drawn in Fig.~\ref{Figsetup}. Note that this corresponds to the external contour of the ejecta curtain. This allows to define  the bottom diameter $D_b$ and the top diameter $D_t$ as the minimal and maximal diameter of the corona.  The height $H$ of the corona is measured as the distance between the mean vertical position of the corona top contour and the granular surface level before impact. As the corona lateral edge appears quite straight except in a small zone at the base of the corona, we extract also the angle $\theta$ formed by the corona edge with the horizontal, by a linear fit of the straight portion.

%% Figure 4 
\begin{figure}
\centerline{ \includegraphics[width=.75\linewidth]{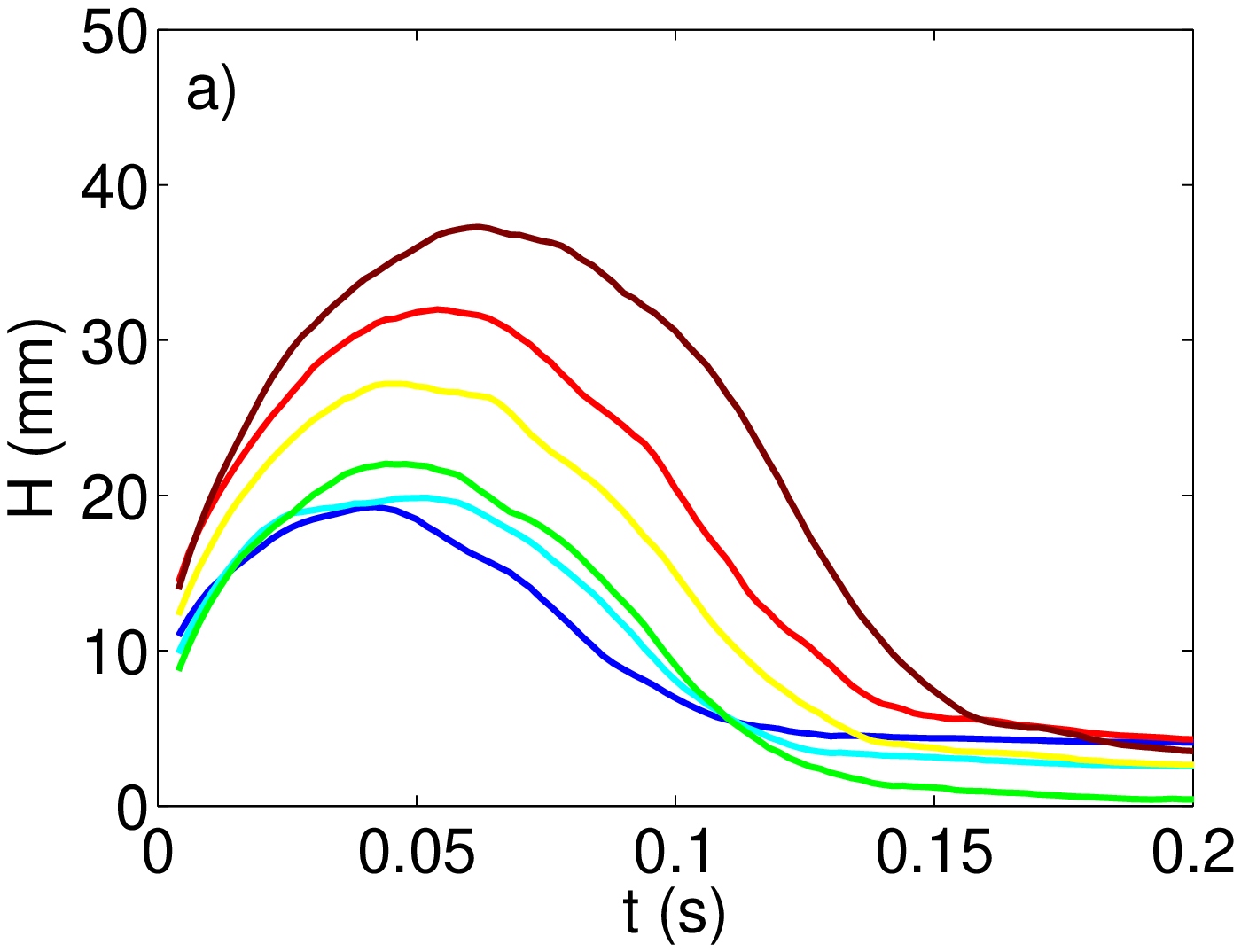} } 
\centerline{ \includegraphics[width=.75\linewidth]{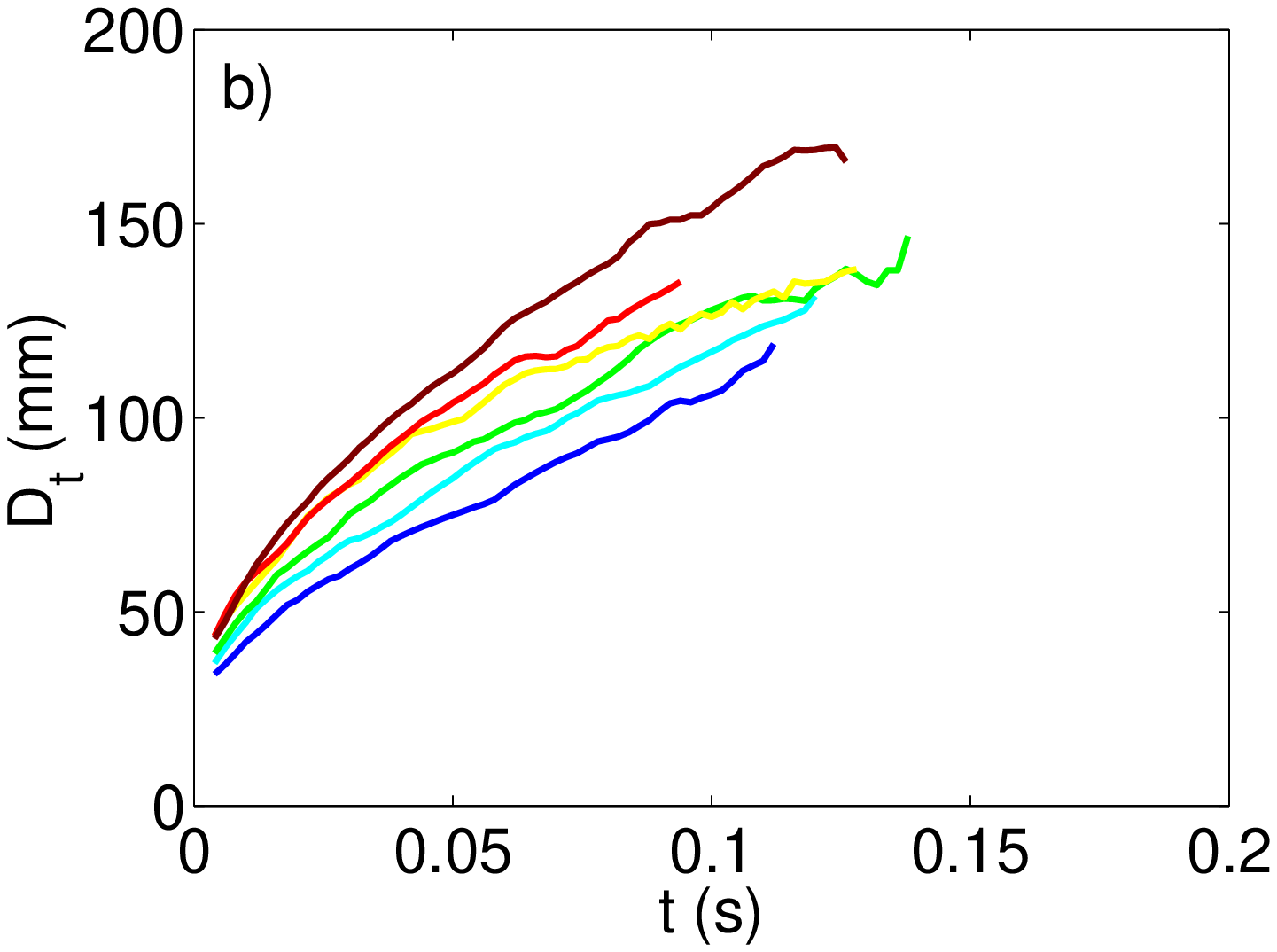} } 
\centerline{ \includegraphics[width=.75\linewidth]{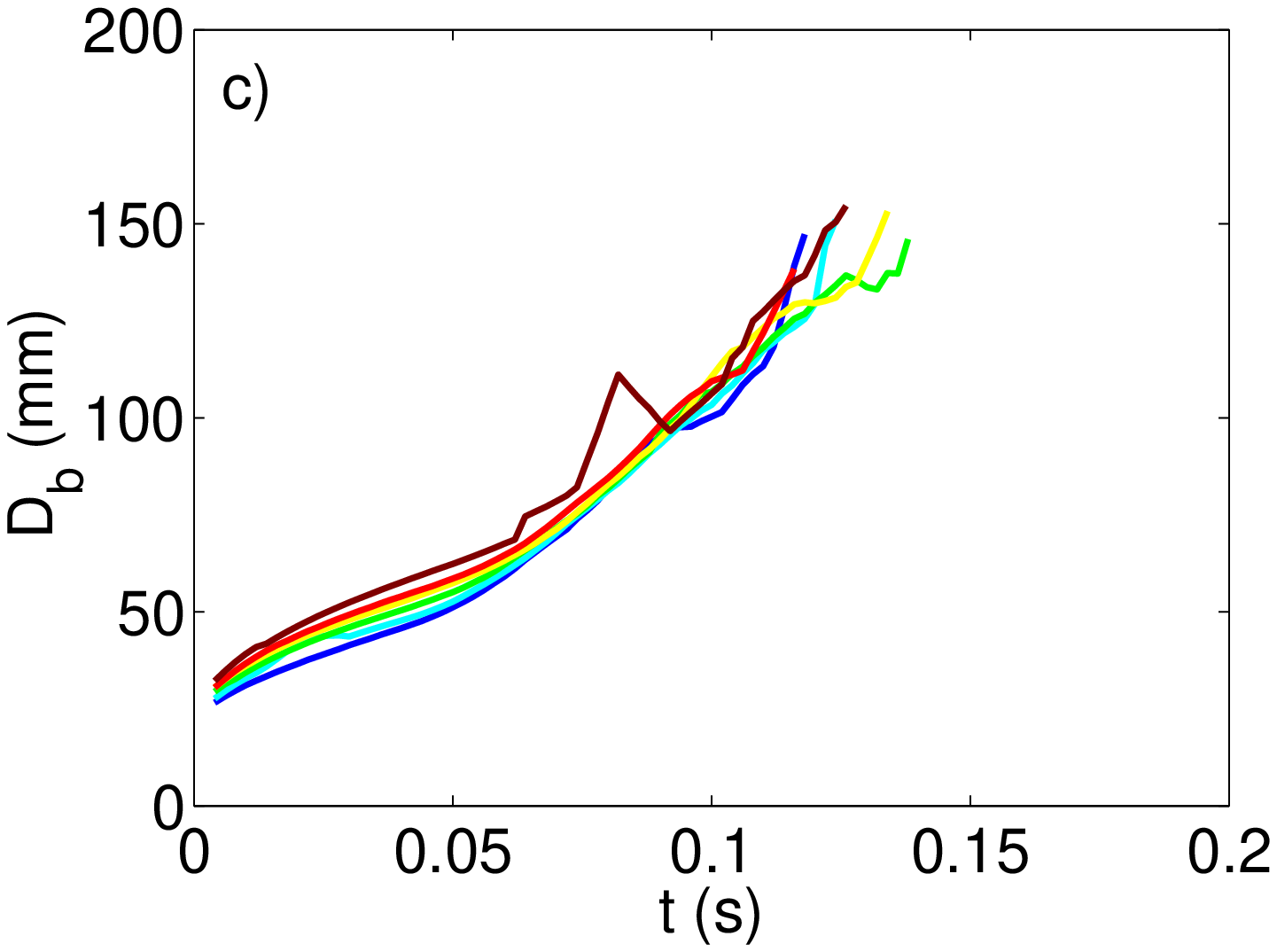} } 
\centerline{ \includegraphics[width=.75\linewidth]{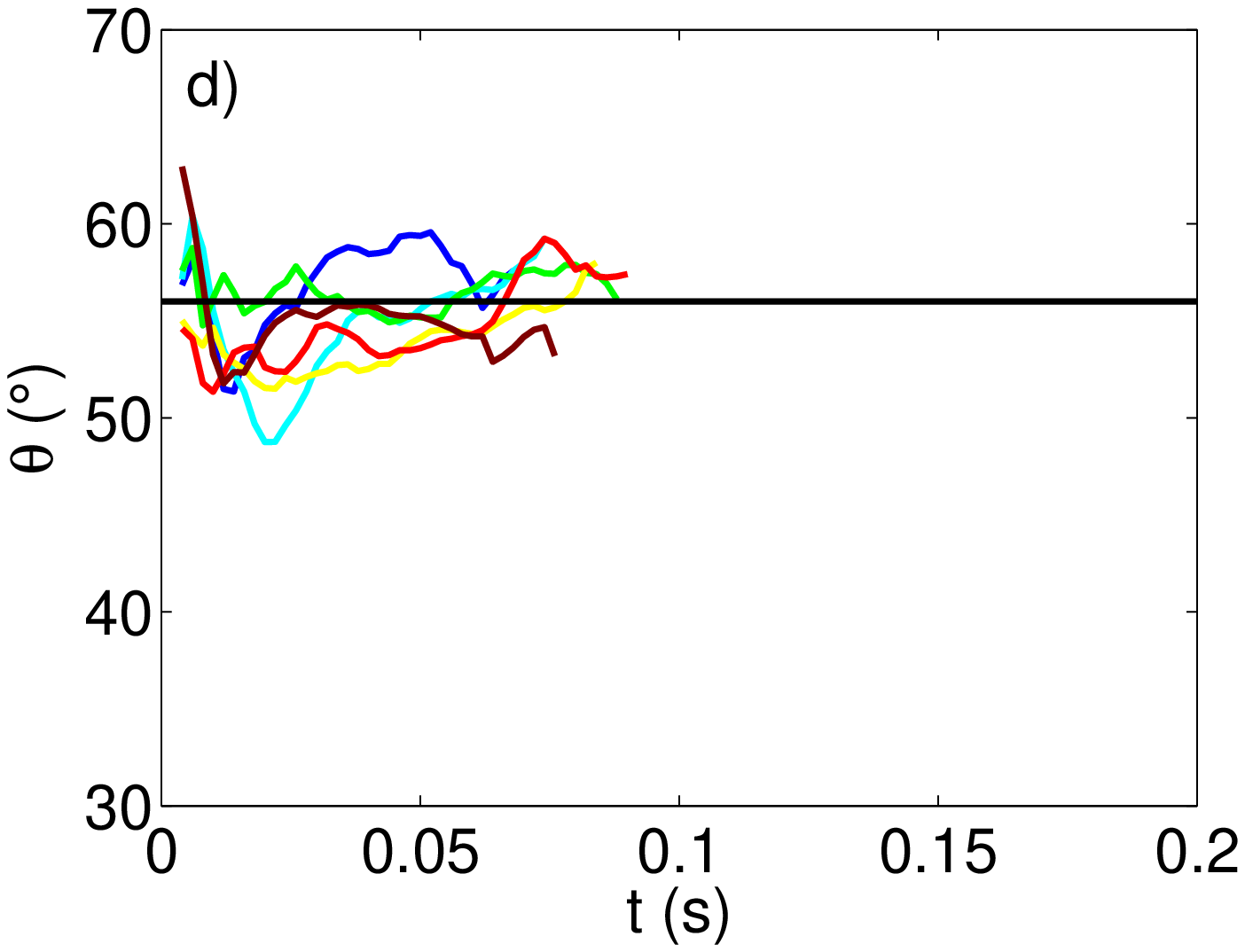} }
\caption{(Color online) Time evolution of a) the height $H$, b) the top diameter $D_t$, c) the bottom diameter $D_b$, and d) the angle of the corona $\theta$ for the same experimental parameters as in Fig.~\ref{Figatot}. The horizontal line in d) is the time and ensemble average of $\theta(t)$ for the $65$ experiments.  }
\label{Figeo}
\end{figure}

The time evolution of the corona, in terms of its height $H$, its top and bottom diameters $D_t$ and $D_b$, and its edge slope $\theta$, is displayed in Fig.~\ref{Figeo} for the same experiments as in  Fig.~\ref{Figatot}. The expansion of the corona is demonstrated  in Fig.~\ref{Figeo}a by the increase of its height $H$ up to a maximal value  denoted $H_{max}$ at time $t_{H_{max}}$ before decreasing. For a given sphere, $H_{max}$ and time $t_{H_{max}}$ increase monotically with the dropping height $h$, so that the different curves of Fig.~\ref{Figeo}a appear in order. Note that for large impact energies (large impact heights $h$), $H$ decreases to a significant non-zero final value, because of the final crater rims lying above the initial free surface~\cite{debruyn03}. 
In the same time, the top diameter $D_t$ and the bottom diameter $D_b$ increase with time, as shown in Fig.~\ref{Figeo}b and c, up to their maximal values, $D_{t\, max}$ and $D_{b\, max}$, 
when the corona disappears and its height vanishes. The evolutions of $D_t$ and $D_b$ are different: About linear for $D_t$ but parabolic for $D_b$. Besides, $D_t$ increases significantly with the dropping height $h$ whereas $D_b$ does not vary so much.
 Note that at the nearly end of the corona life, for vanishing height $H$,  the values of $D_t$ and $D_b$ become very noisy and are thus not shown here.
The angle  $\theta$ the corona edge forms with the horizontal, shown in Fig.~\ref{Figeo}d,  is roughly constant as a function of time and whatever the dropping height $h$, and equal to about $56^{\circ}$ with relative variations  of $\sim5\%$.
The same kind of evolutions of all these parameters are observed for the different tested spheres.

The corona evolution reported here for granular impacts can be compared to the corona evolution for the liquid case~\cite{yarin}. For the liquid case, the corona base spreads radially as the square root of time whereas it seems more close to a quadratic evolution for the granular case (see $D_b(t)$ in Fig.~\ref{Figeo}c). The scaling laws for spreading is thus different in the granular and liquid cases, with inverse curvatures in the $D_b(t)$ plot. Besides, the angle of the granular corona is found independent of the impact velocities, as found for the liquid corona, with a value closer to the case of shallow liquid layers than the case of deep liquid layers~\cite{fedorchenko}. This may be related to the collisional chains that redirected the impact velocity within a few grains layer only~\cite{crassous}.

%% Figure 5
\begin{figure}
\centerline{ \includegraphics[width=.75\linewidth]{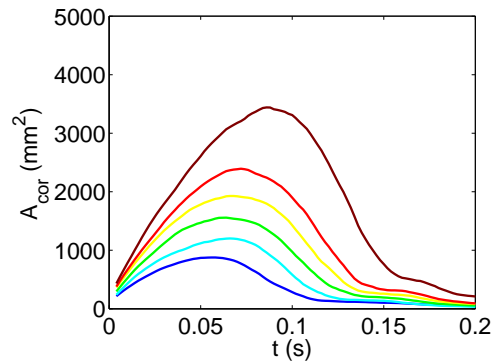} }
\caption{(Color online) Time evolution of the apparent surface area of the corona $A_{cor}$ for the same experimental parameters as in Fig.~\ref{Figatot}.}
\label{Figacor}
\end{figure}
%%

%% Figure 6
\begin{figure}
\centerline{ \includegraphics[width=.75\linewidth]{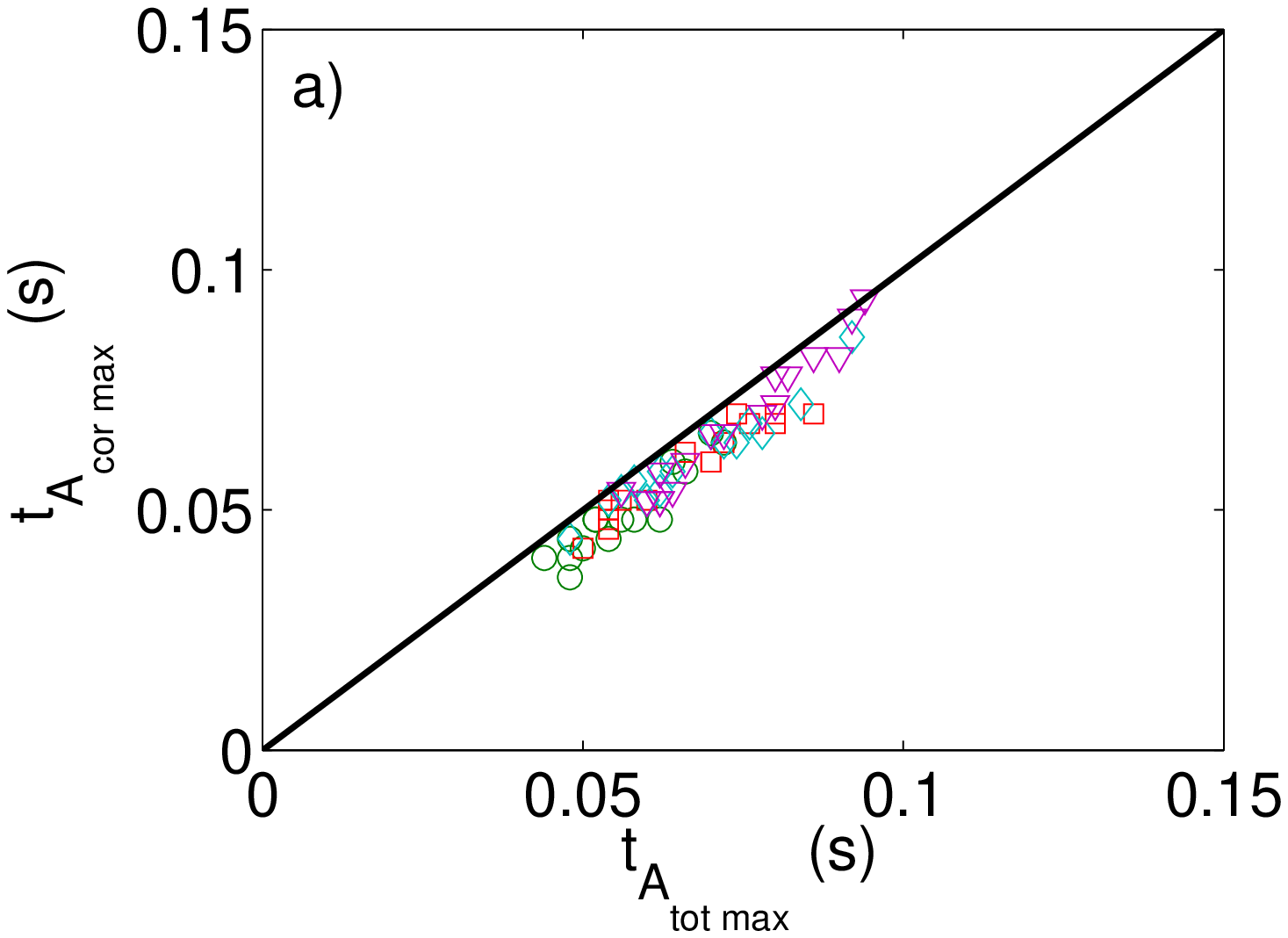} } 
\centerline{ \includegraphics[width=.75\linewidth]{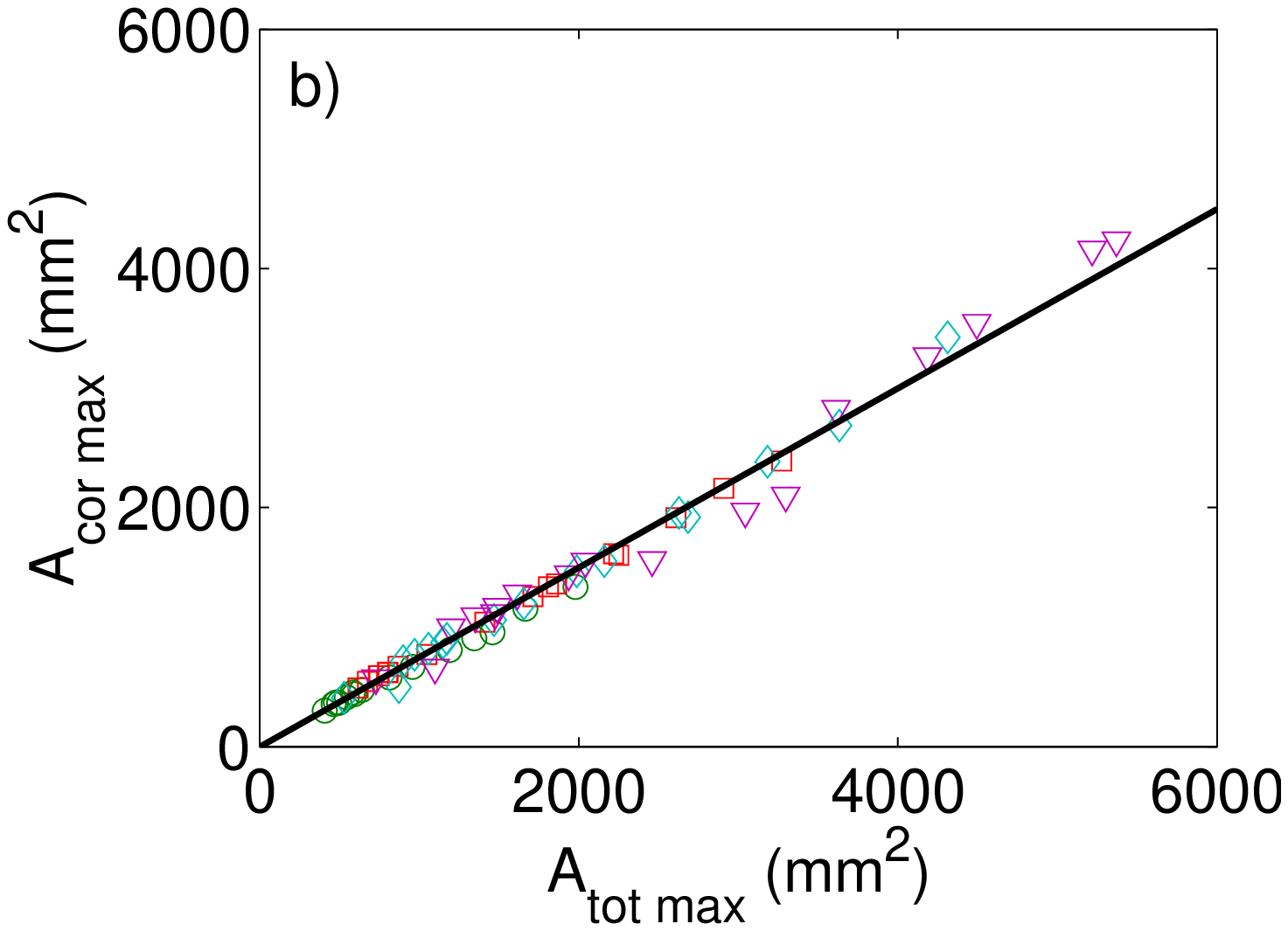} } 
\caption{(Color online) a) Time $t_{A_{cor\, max}}$ for maximal value of the corona apparent area  $A_{cor}$ as a function of time $t_{A_{tot\, max}}$ for maximal value of the total apparent area $A_{tot}$. The solid line is of slope $1$. b) Maximal values  $A_{cor\, max}$ of $A_{cor}$ as a function of maximal values $A_{tot\, max}$ of  $A_{tot}$. The solid line is a linear fit. Data symbols are for the 65 experiments with different dropping heights $h$ for impacting spheres of radius $R$ = 5.15 ($\circ$), 6.75 ($\square$), 7.55 ($\diamond$) and 9.50 ($\triangledown$)~mm. }
\label{Figcompa}
\end{figure}

With the measurements of the geometrical parameters of the corona, we can now define the apparent area of the corona $A_{cor}$, that is related to the amount of ejected grains contained in the corona, and compare it to the total apparent area of ejected grains $A_{tot}$, related to the total amount of the ejected grains, both in the corona and isolated. $A_{cor}$ is measured as the area included inside the corona contour, which is not far from the area $H(D_t+D_b)/2$ corresponding to the approximate corona trapezium shape in the images. The time evolution of $A_{cor}$ is reported in Fig.~\ref{Figacor} for the same experimental parameters as in Figs. 2 and 4. Even if $A_{cor}$ is always smaller than $A_{tot}$, the evolution of $A_{cor}$ is qualitatively the same as the one of $A_{tot}$ (Fig.~\ref{Figatot}). This is confirmed by the quantitative comparison of the corresponding coordinates of the curve maxima, ($t_{A_{tot\, max}}$, $A_{tot\,max}$) and ($t_{A_{cor\, max}}$, $A_{cor\,max}$), that are reported in Fig.~\ref{Figcompa} for all the experiments (all dropping heights, all different impacting spheres). The evolutions of $A_{cor}$ and $A_{tot}$ are synchronized in time as illustrated by the equality $t_{A_{cor\, max}}\simeq t_{A_{tot\, max}}$ (Fig.~\ref{Figcompa}a). Besides, $A_{cor\,max}$ is found proportional to $A_{tot\,max}$ with the same ratio for all experiments: $A_{cor\, max}\simeq0.75A_{tot\, max}$ (Fig.~\ref{Figcompa}b). All this suggests that the investigation of the corona dynamics is a good first order for the study of the dynamics of grain ejection due to an impact. Let us now interpret these experimental results by a simple ballistic model.  

\section{Ballistic model for grain ejection} \label{sec_mod}
The dynamics of the grains is found axisymmetric  and rapidly contact forces between grains play  no role (Fig.~\ref{Figseq}). Furthermore air friction can be estimated and is negligible compared to the grain weight. Thus rapidly after impact, each grain trajectory corresponds to a parabolic free flight under the action of gravity alone. Using these preliminary remarks, we now build a very simple axisymmetric model that reproduces the observed corona dynamics in the $(r,z)$ plane of cylindrical coordinates. 

\begin{figure}
\centerline{ \includegraphics[width=0.75\linewidth]{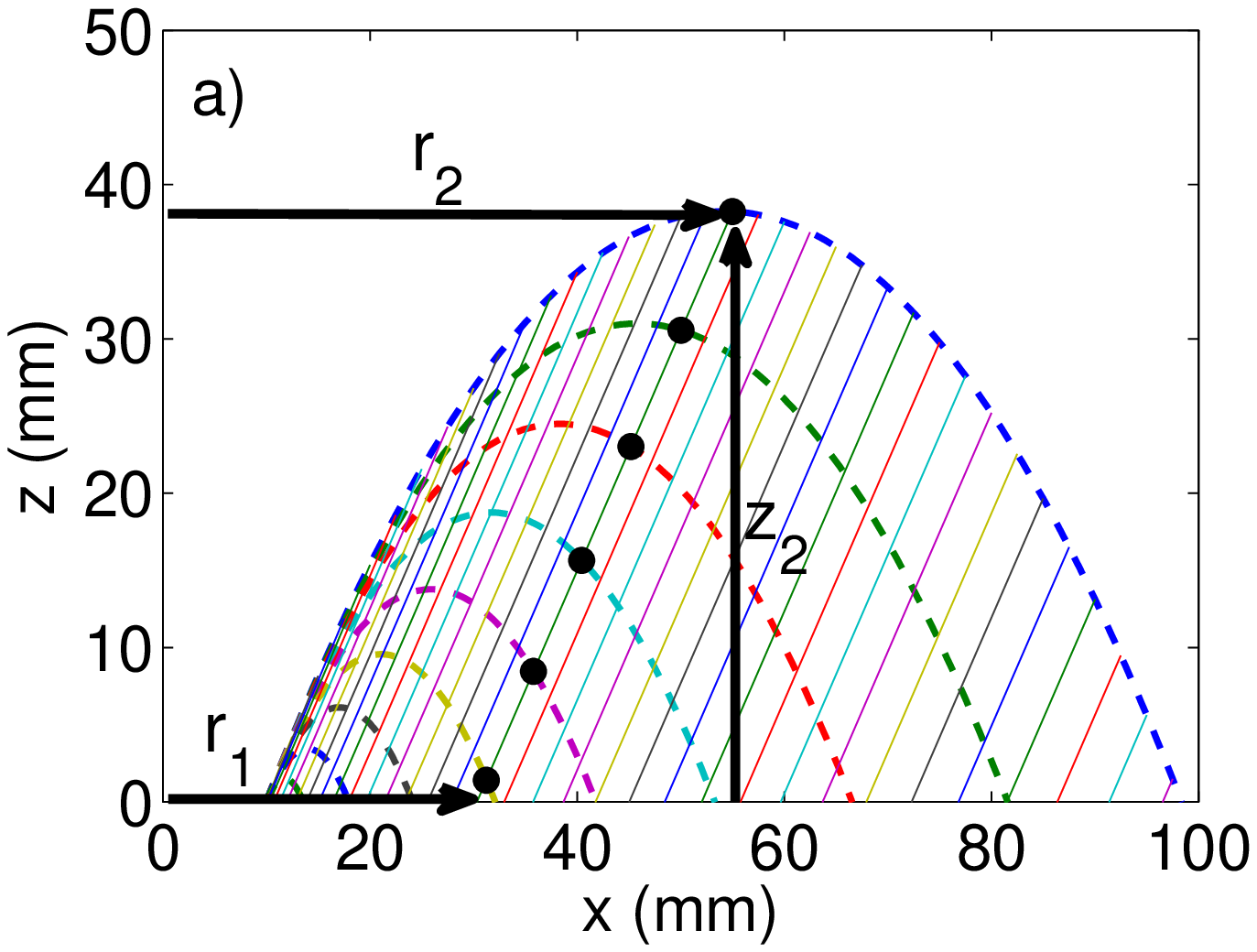} }
\centerline{ \includegraphics[width=0.75\linewidth]{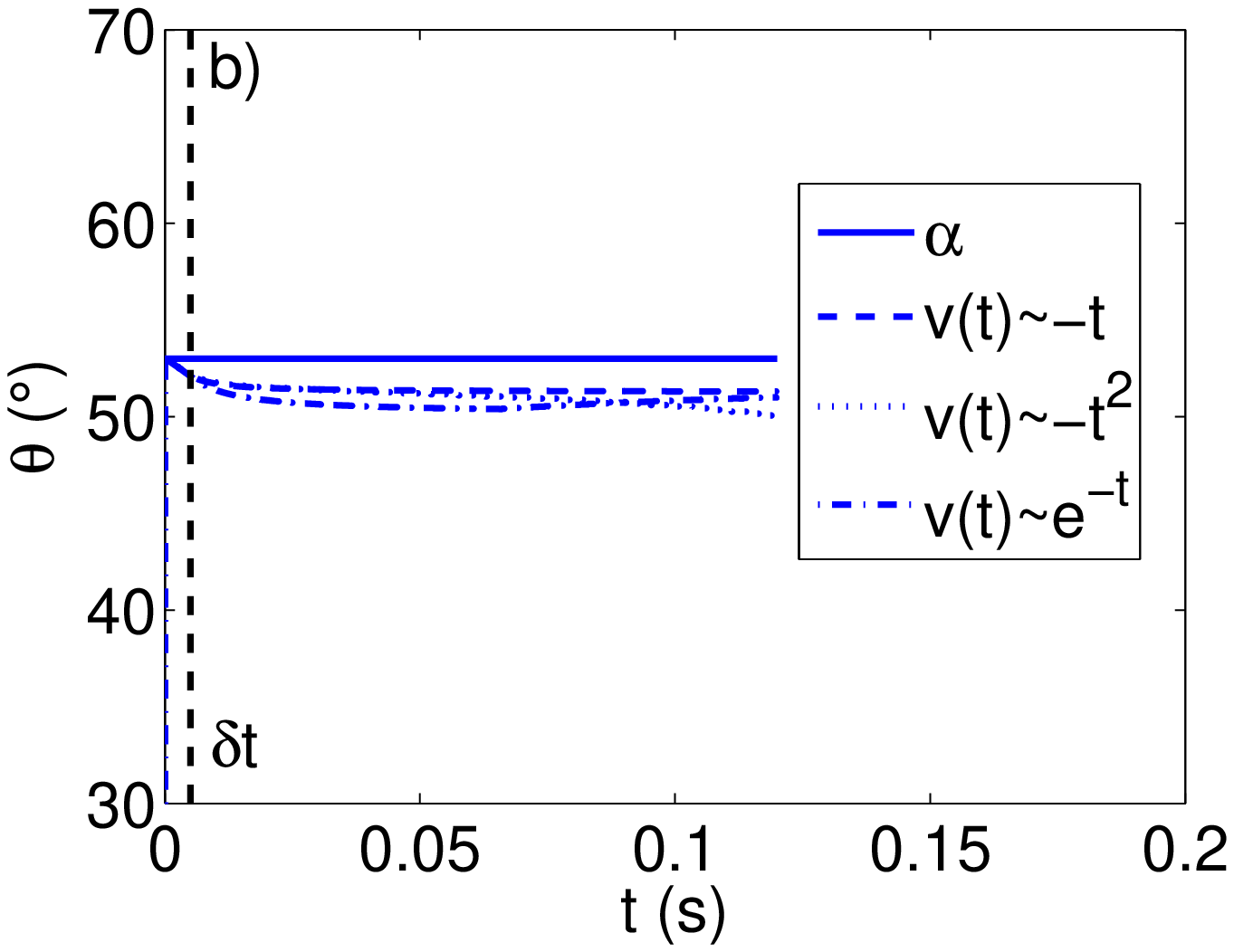} }
\caption{(Color online) a) Ballistic model for grains ejected instantaneously from the same position $r_0$ in the same direction making the angle $\alpha$ with the horizontal, with different velocity amplitudes. Trajectories are drawn in dashed lines. At any time grains align along a straight line forming the  corona edge of angle $\theta=\alpha$ drawn in continuous lines. b) Time evolution of the angle $\theta$ of the cone edge  if grains are ejected during a finite time $\delta t=5$~ms in the direction $\alpha=53^{\circ}$, with a linear, quadratic or exponential decrease of ejection velocity from the initial value $u_0=1$~m/s.}
\label{FigBalistic}
\end{figure}

As a first attempt, we assume that all grains start at the same time $t=0$, from the same position $(r_0,0)$, in the same initial direction making an angle $\alpha$ with the horizontal, but with different velocity amplitudes. It is thus easy to show (Fig.~\ref{FigBalistic}a) that at any time $t$ all the grains will be located on a cone making the same angle $\alpha$ with a bottom diameter $2r_1$ increasing quadratically with time as $2r_1(t)=2r_0+gt^2/\tan \alpha$. Furthermore, if $u_0$ is the maximal amplitude of initial velocity, the top diameter of the cone $2r_2$  increases linearly with time as $2r_2(t)=2r_0+2u_0\cos \alpha\, t$ and the height of the cone $z_2$ evolves  quadratically  in time as $z_2(t)=u_0\sin \alpha\, t-gt^2/2$. 
These analytic results are quite similar to the dynamics observed in Fig.~\ref{Figseq} and to the time evolution of the corona parameters $H$, $D_t$, $D_b$ and $\theta$ presented in Fig.~\ref{Figeo}.
The hypothesis of an instantaneous release of all the grains at the same place and the same time $t = 0$ is however clearly oversimplified. Indeed grains are ejected during a time interval $\delta t$ of the order of the penetration time $R/U_c \sim 5$ ms and, as the projectile decelerates the ejection velocity of the grains should be a decreasing function of time. In Fig.~\ref{FigBalistic}b, we plot the apparent cone angle $\theta(t)$ of the moving grains  when released with the same ejection angle $\alpha$ but with various models of time decreasing velocities $u_0(t)$ during the time interval $\delta t=5$~ms: A linear decrease [$u_{ej}(t)=u_0(1-t/\delta t)$], a quadratic decrease [$u_{ej}(t)=u_0(1-t^2/\delta t^2)$], or an exponential decrease [$u_{ej}(t)=u_0\exp(-t/\delta t)$] from  the initial value $u_0=1$~m/s. Except at very short times the edge angle $\theta$ remains constant and always close to the ejection angle $\alpha$. While the time evolutions for the top diameter $2r_2$ and the height $z_2$ of the apparent cone are obviously the same than for an instantaneous release, it appears to be also the same but with a small delay time $\delta t$ for the bottom diameter $2r_1$. As this delayed  grain ejection and initial velocity decrease have no visible effect in the corona dynamics, we will use in the following the first simpler model where all the grains are ejected {\sl at the same time} with ejection velocities $0\leq u_{ej} \leq u_0$.

%% Figure 7

In the experiments we measured the time evolution of the corona parameters, i.e. the optically opaque zone formed by ejected grains, when viewed from the side. The relation between this  corona that corresponds to a 2D projection of the 3D real dynamics with this latter is not straightforward. But  the bottom diameter $D_b$ and the edge angle $\theta$ are clearly the same. Furthermore, Fig.~\ref{Figcompa} shows that most of the grains are inside the corona.  In the following we will then assume that the height $H$ and the top diameter $D_t$ of the corona in the experiments are identical to the height $z_2$ and the top diameter  $2r_2$ of the cone in the ballistic model.
Assuming this, the time evolution of $H$, $D_t$, $D_b$ and $\theta$ are given by the following set of equations:
\begin{subequations} \label{eq_modbal}
 \begin{eqnarray}
  H(t)&=&u_0\sin \alpha\, t -gt^2/2 + H(0) \mathrm{ , } \label{eq_h} \\
  D_t(t)&=&D_t(0)+2u_0\cos \alpha\, t \mathrm{ , } \label{eq_dh} \\
  D_b(t)&=&D_b(0)+gt^2/\tan \alpha \mathrm{ , } \label{eq_d0} \\
  \theta &=& \alpha  \mathrm{ , }
 \end{eqnarray}
\end{subequations}
An initial height $H(0)$ has been introduced in the dynamics of $H(t)$ as we observe in the experiment an initial swelling of the granular bed before any grain ejection. This swelling corresponds to a deformation of the substrate when the projectile starts to penetrate the bed. This deformation process is clearly not contained in our ballistic description.
The horizontal position of ejection $r_0$ is related to the projectile size: We checked that $D_t(0)$ and $D_b(0)$ increase with the impacting sphere diameter $2R$. 
The set of equations~(\ref{eq_modbal}a)-(\ref{eq_modbal}c) can be written in dimensionless form as:
\begin{subequations} \label{eq_modbaladim}
 \begin{eqnarray}
  \frac{H(t)-H(0)}{H_{max}-H(0)}&=&2\left(\frac{t}{t_{H_{max}}}\right)-\left(\frac{t}{t_{H_{max}}}\right)^2 \mathrm{ , } \label{eq_hrescal} \\
  \frac{D_t(t)}{D_t(0)}&=&1+\frac{2u_0\cos \alpha}{D_t(0)} t \mathrm{ , } \label{eq_dhrescal} \\ 
  \frac{D_b(t)}{D_b(0)}&=&1+\frac{g}{D_b(0)\tan\alpha} t^2 \mathrm{ . } \label{eq_d0rescal}   
 \end{eqnarray}
\end{subequations}
with the two scaling parameters:  $t_{H_{max}}=u_0\sin \alpha/g$ and $H_{max}-H(0)= g t_{H_{max}}^2/2$.

\begin{figure}
\centerline{ \includegraphics[width=.75\linewidth]{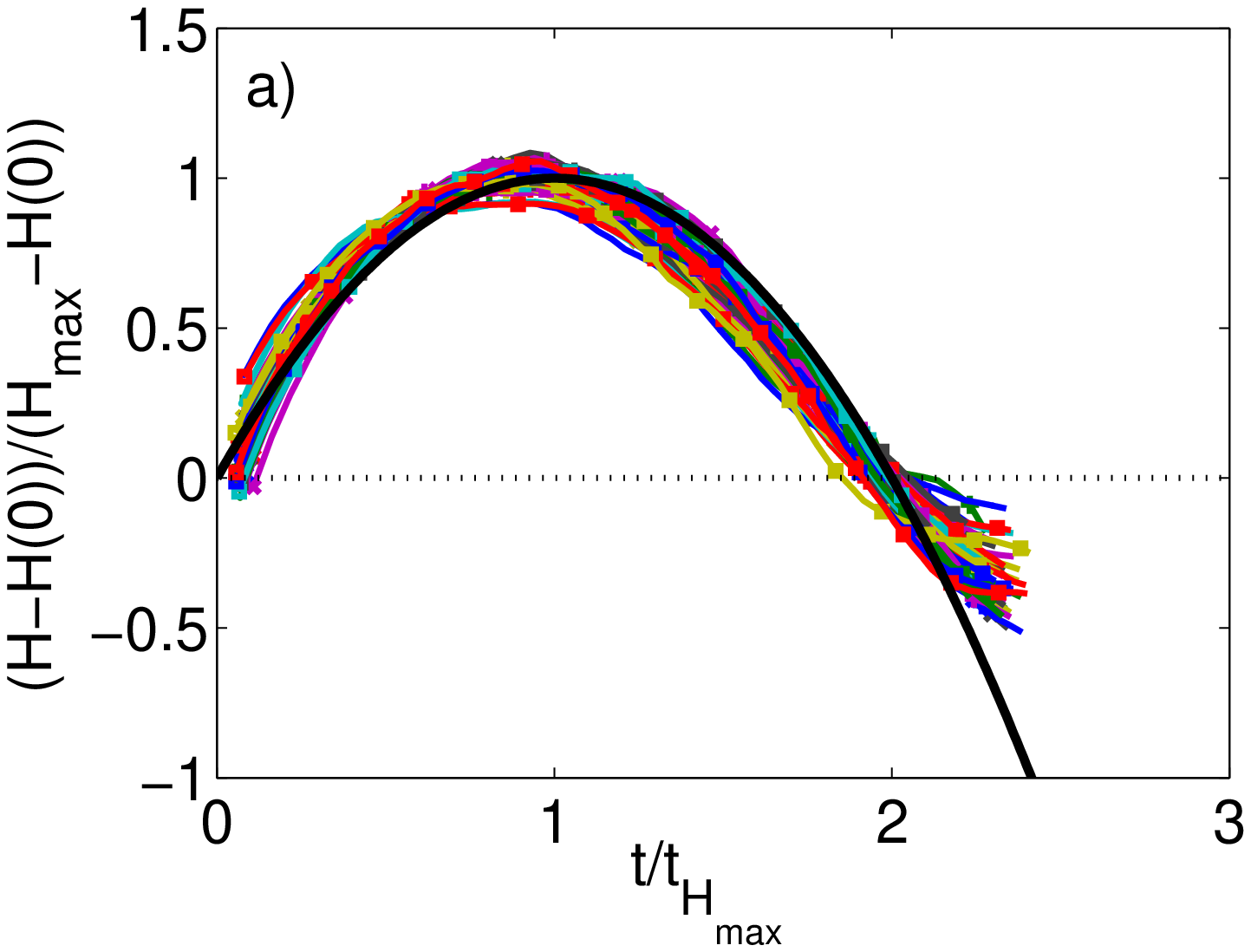} } 
\centerline{ \includegraphics[width=.75\linewidth]{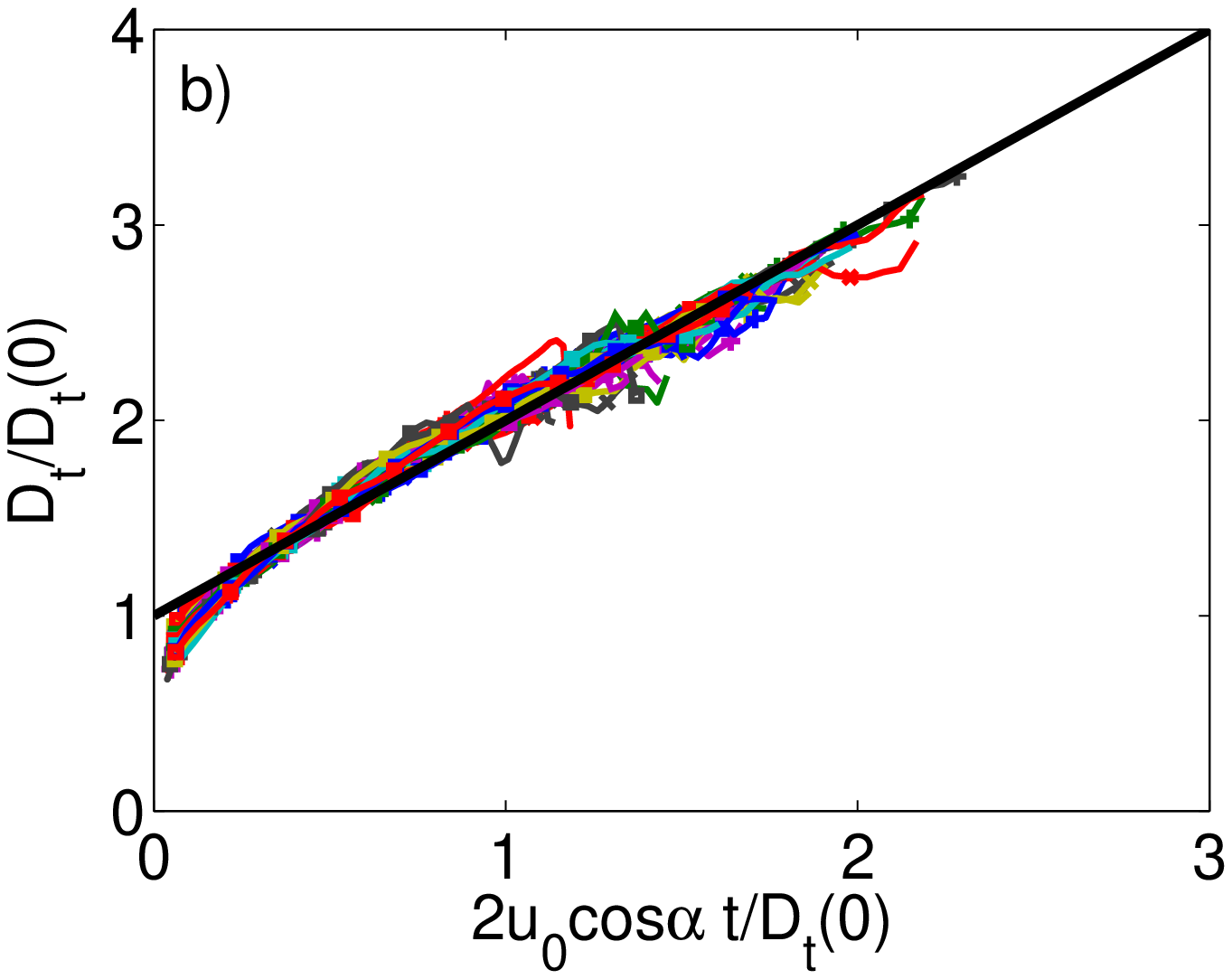} }
\centerline{ \includegraphics[width=.75\linewidth]{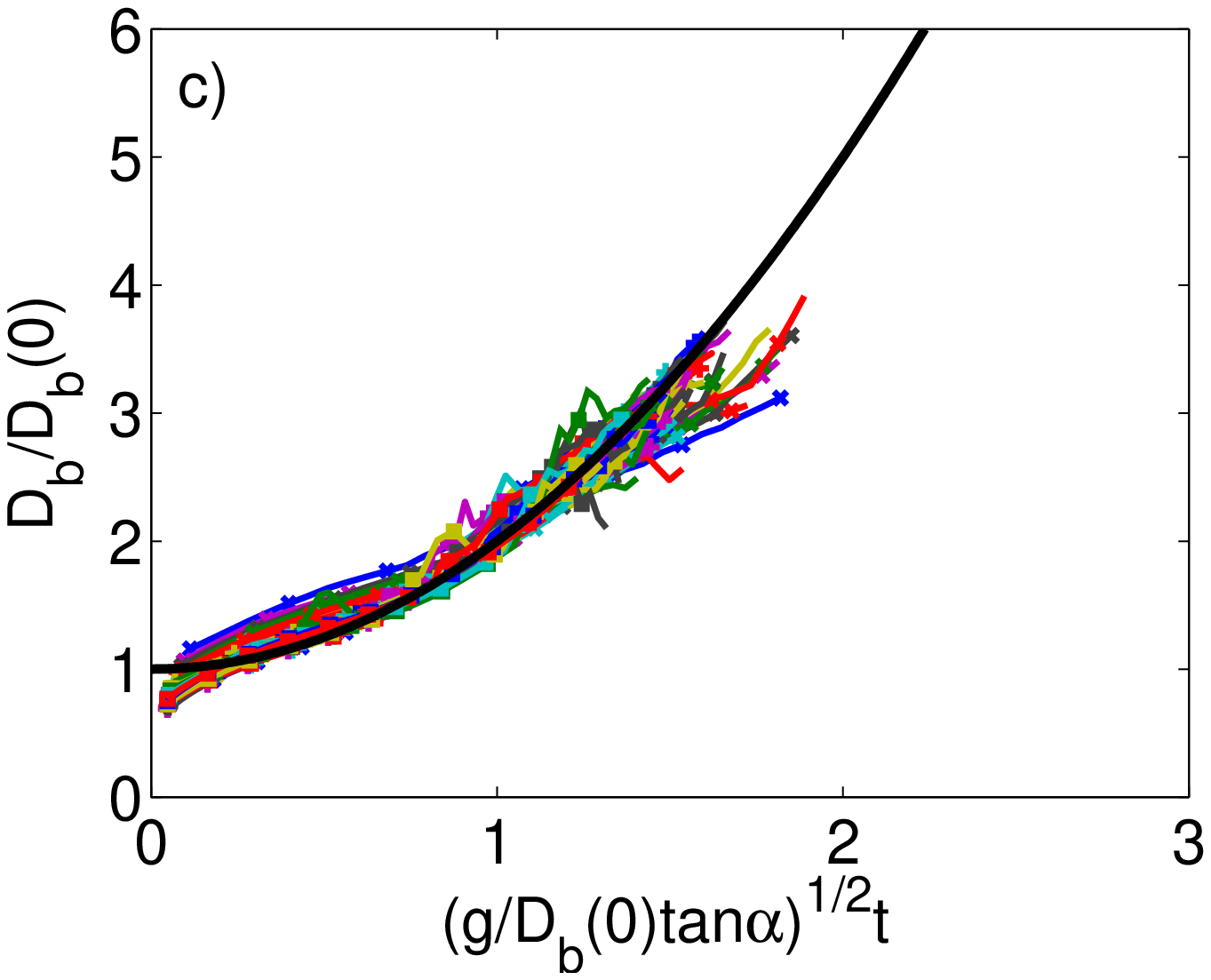} }
\caption{(Color online) Rescaled data of a) heights $(H-H(0))/(H_{max}-H(0))$, b) top diameter $D_t/D_t(0)$  and c) bottom diameter $D_b/D_b(0)$ as a function of rescaled times appearing in Eqs.~\ref{eq_modbaladim}  for all of the $65$ experiments.  }
\label{FigRescaled}
\end{figure}

For each of the 65 experiments, our experimental data for the height $H(t)$ and the top diameter $D_t(t)$ of the corona are well fitted by the ballistic model equations as shown in the non-dimensional plots of  Fig.~\ref{FigRescaled}a and b. The evolution of $H$ is basically quadratic, while $D_t$ is rather linear in time. The fitting parameters $u_{0} \sin \alpha$ for $H(t)$ and $u_{0} \cos \alpha$ for $D_t$ give access to the typical grain ejection velocity $u_0$ and to the typical ejection angle $\alpha$ for each impact experiment. 
The time evolution of the bottom diameter of the corona $D_b(t)$ is shown in the non-dimensional plot of Fig.~\ref{FigRescaled}c, with the already deduced fitting parameters: The predicted quadratic increase of $D_b$ is rather well followed by the experimental data.

%% Figure 8
%

%% Figure 9
\begin{figure}
\centerline{ \includegraphics[width=.75\linewidth]{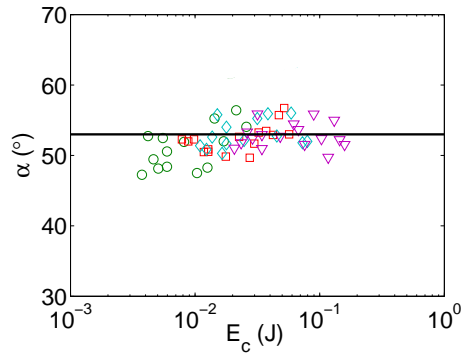} }
\caption{(Color online) Fitted ejection angle $\alpha$ as a function of the impact energy $E_c$ for the $65$ experiments (same symbols as in  Fig.~\ref{Figcompa}).  }
\label{FigAngle_ejection}
\end{figure}

The angle of ejection $\alpha$, extracted from the fits of Fig.~\ref{FigRescaled},  is found to be constant for all experiments: $\alpha\simeq53^{\circ}\pm3^{\circ}$, which corresponds well to the corona edge angle measured previously (Fig.~\ref{Figeo}d): $\theta \simeq 56^{\circ} \pm 3^{\circ}$.  We have checked that  $\alpha$ and $\theta$ are robustly constant for all experiments and do not depend on any impact parameter, as shown in Fig.~\ref{FigAngle_ejection}: No clear correlation of $\alpha$ with the impact energy $E_c$ can be seen. 
The equality found between the fitted ejection angle $\alpha$ and the measured corona edge angle $\theta$ is again in favour of the present grain ejection scenario.

From eqs.~(\ref{eq_modbal}a)-(\ref{eq_modbal}c),  we can calculate the area of the corona  $A_{cor}(t)$: The predicted evolution is in good agreement with experimental results and  gives $t_{A_{cor\, max}} = 1.36 t_{H_{max}}$ very close to the coefficient $1.38$ measured experimentally.

%% Figure 10
\begin{figure}
\centerline{ \includegraphics[width=.75\linewidth]{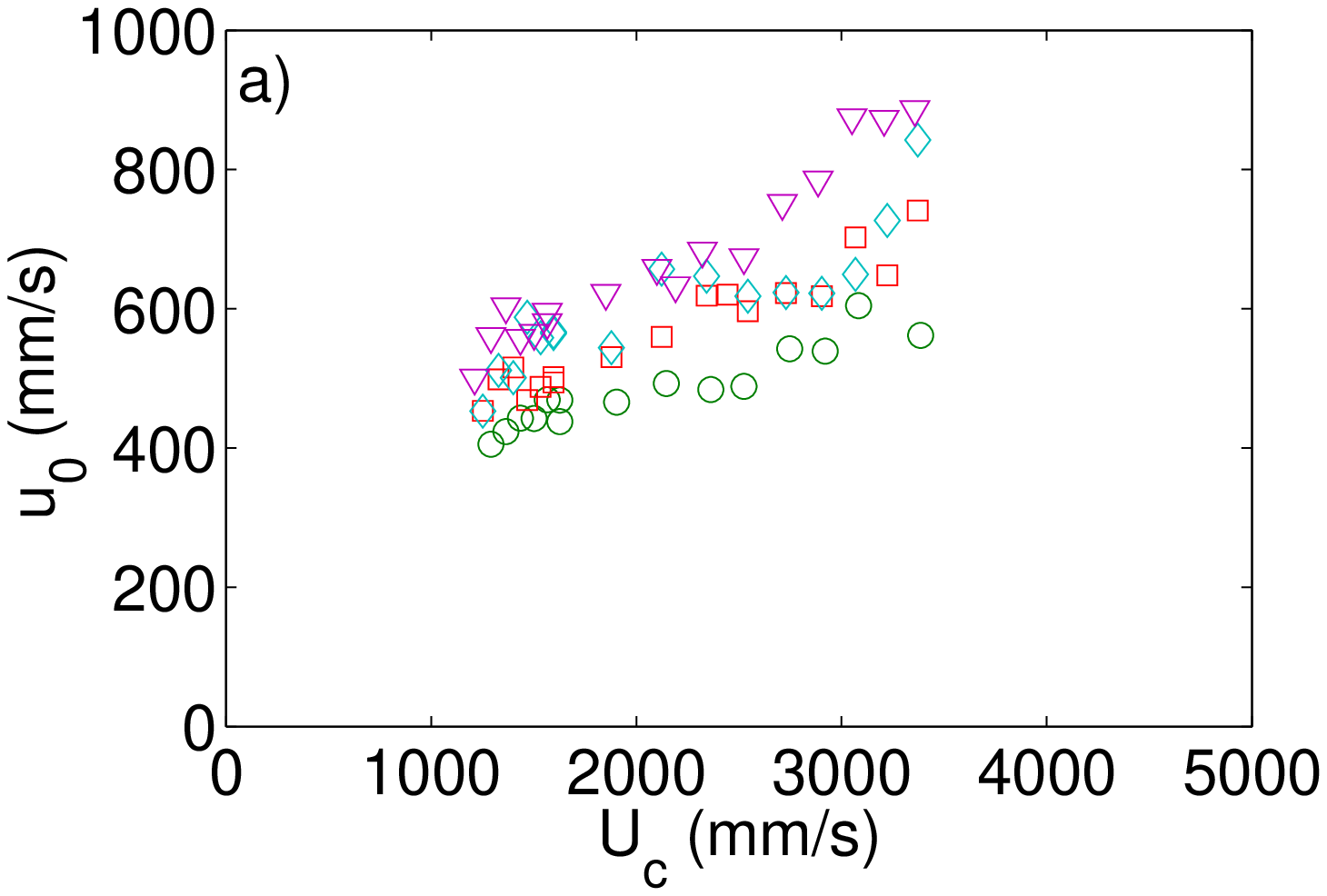} }
\centerline{ \includegraphics[width=.75\linewidth]{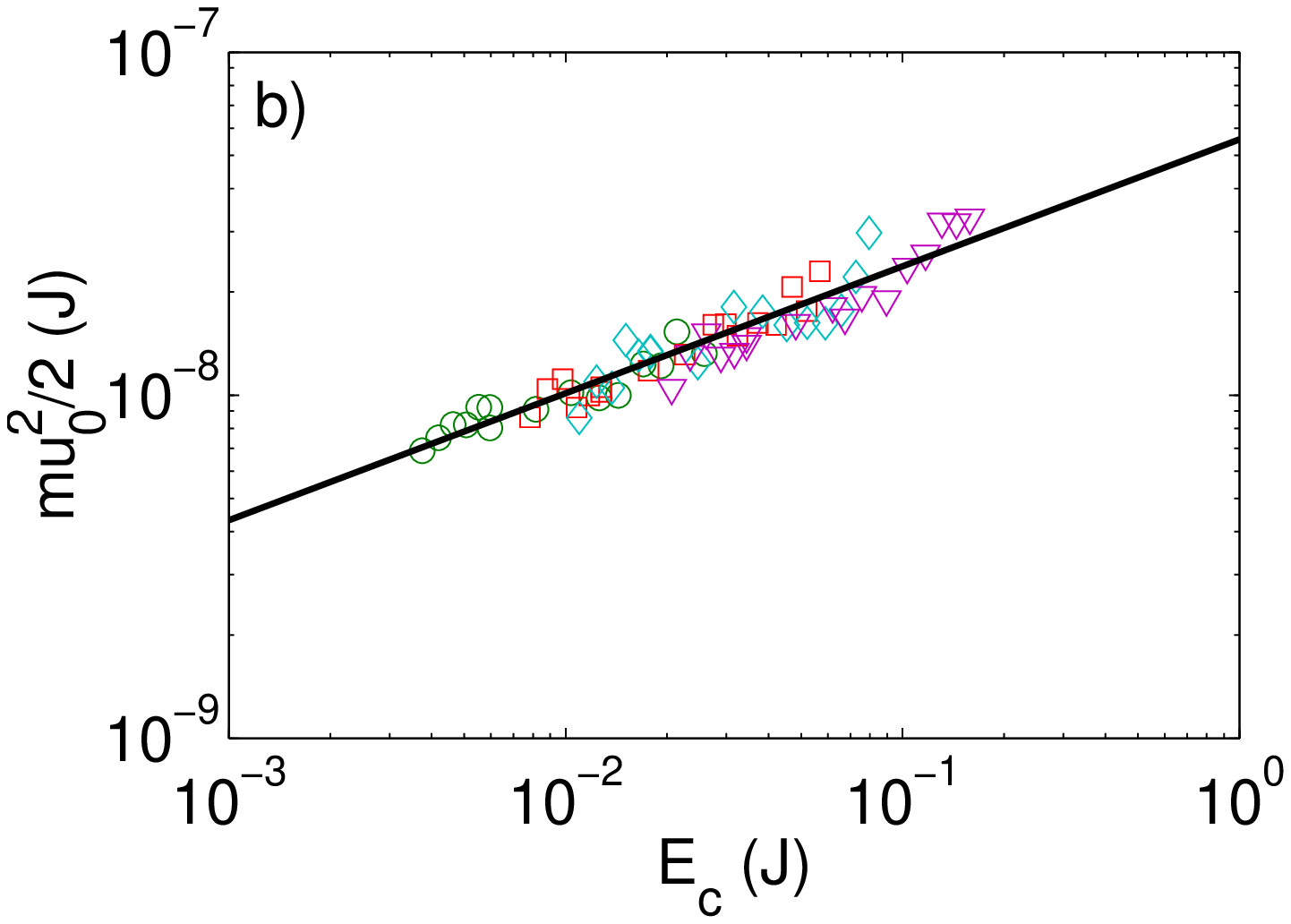} }
\caption{(Color online) Variations of ejection parameters  with the impact parameters: a) Ejection velocity $u_0$ as a function of impact velocity $U_c$; b) Kinetic energy of one grain $mu^2_0/2$ as a function of impact energy $E_c$ and the best power law fit Eq.~(\ref{eq_v0ec}).  Same symbols as in Fig.~\ref{Figcompa}.}
\label{FigEjectionParameter}
\end{figure}
%%

%% Figure 11
\begin{figure}
\centerline{ \includegraphics[width=.75\linewidth]{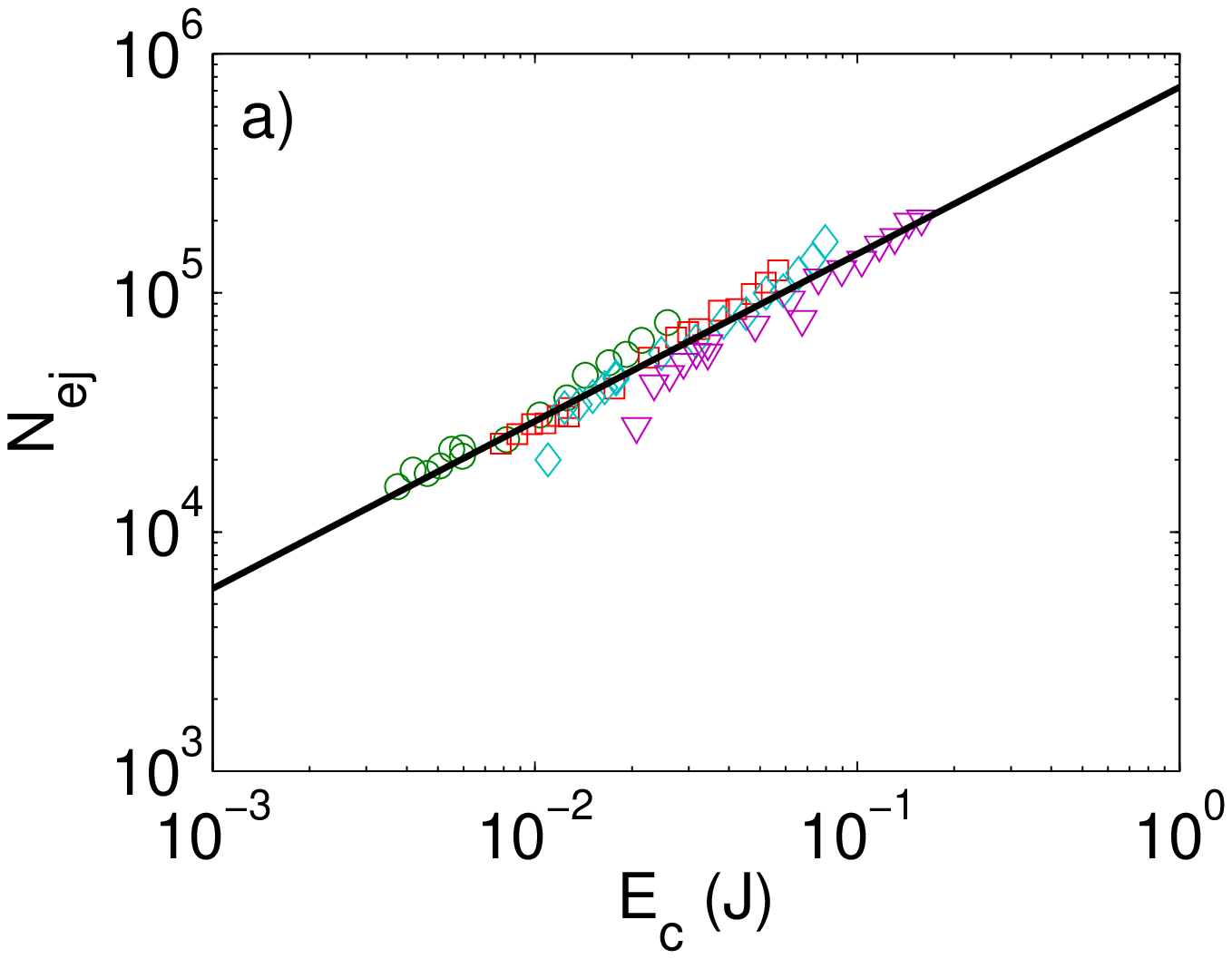} }
\centerline{ \includegraphics[width=.75\linewidth]{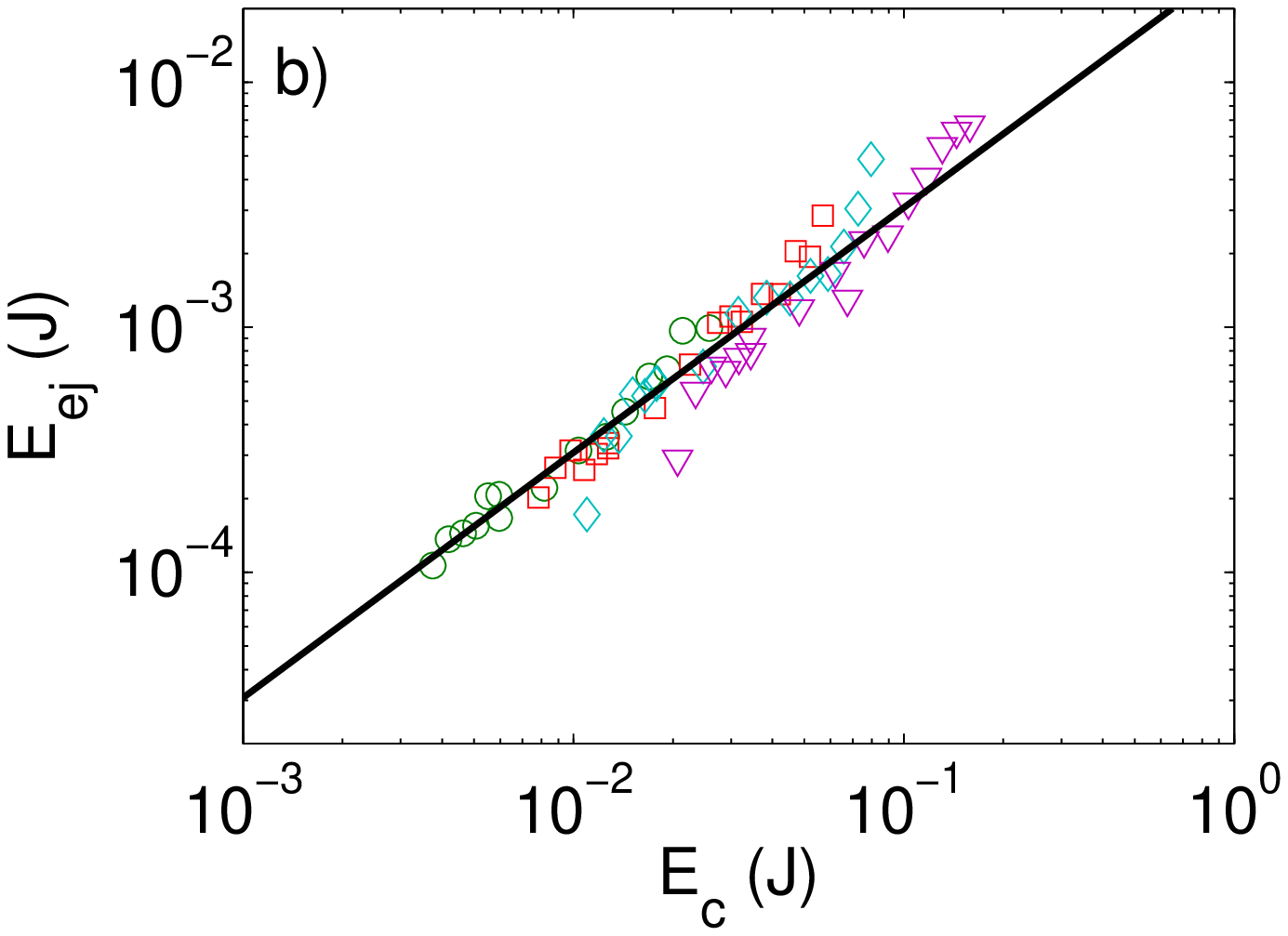} }
\caption{(Color online)  a) Number of ejected grains $N_{ej}$ as a function of impact energy $E_c$  for all experiments, and  best power law fit given in Eq.~(\ref{eq_nejec}). b) Kinetic energy of ejecta $E_{ej}$ as a function of impact energy $E_c$ and  best power law fit given in  Eq.~(\ref{eq_enejec}).  Same symbols as in Fig.~\ref{Figcompa}.}
\label{FigEnergy}
\end{figure}

Although we do not measure directly the grain ejection velocities, the fits of  Fig.~\ref{FigRescaled} give a characteristic ejection velocity $u_0$.  
Figure~\ref{FigEjectionParameter}a shows the slow increase of $u_0$ with the impact velocity $U_c$, and its dependence on the sphere radius, where we can see that $U_c$ is not the relevant parameter to account for the variations of $u_0$. 
The kinetic impact energy $E_c = MU_c^2/2$ is the relevant parameter, as  shown in Fig.~\ref{FigEjectionParameter}b where the kinetic energy of one ejected grain $m u_0^2/2$ is plotted as a function of the impact energy $E_c$ in  log-log axes. The best power law fit is:

\begin{equation}
 m u_0^2/2 \simeq 0.56 \,10^{-9} E_c^{0.37\pm0.05} \mathrm{ , }
 \label{eq_v0ec} 
\end{equation}
showing a moderate effect of the impact energy on the kinetic energy of one ejected grain.
From the ballistic model presented above and this experimental result of the scaling of the ejection velocity $u_0$ with the impact energy $E_c$, we can infer the scaling for the total duration of the corona ballistic dynamics  $T$,  for its maximal height $H_{max}$ and its maximal top diameter $D_{t\,max}$ with the impact energy $E_c$ as:  
\begin{eqnarray}
T = 2t_{H_{max}} & = & 2u_0\sin\alpha/g \propto  E_c^{0.19} \mathrm{ , }  \\ 
H_{max} & = & gt^2_{H_{max}}/2 \propto  E_c^{0.37} \mathrm{ , }  \\ 
D_{t\,max} & = & u^2_0\sin2\alpha/g \propto  E_c^{0.37} \mathrm{ , }  
\end{eqnarray}
allowing to characterize the duration and  the extension of deposits upon an impact. These results show that the corona duration $T$ depends only slightly  on the impact energy $E_c$, whereas the corona expansion depends moderatly on $E_c$.   

In Eq.~(\ref{eq_nej}), the characteristic number of ejected grains $N_{ej}$ has been related to  the maximal value of the total apparent surface of all of the ejected grains $A_{tot\,max}$, through the unknown product $\overline{n} w$. 
We will estimate in  section~\ref{sec_disc} that $\overline{n}w\simeq 12~mm^{-2}$ and we use from now this value to calculate the number of ejected grains $N_{ej}$ according to Eq.~(\ref{eq_nej}).  In Fig.~\ref{FigEnergy}a $N_{ej}$ is plotted as a function of the impact energy $E_c$  in log-log axes. We see that whatever the projectile size all the experiments collapse close to the power law:  
\begin{eqnarray}
N_{ej}\simeq7.2\,10^{5} E_c^{0.70\pm0.05} \mathrm{ . } \label{eq_nejec} 
\end{eqnarray}

We can now estimate the kinetic energy  transmitted to all the ejected grains as $E_{ej}\simeq N_{ej}mu_0^{2}/2$.  
Figure~\ref{FigEnergy}b shows the evolution of $E_{ej}$ as a function of $E_c$.  The total kinetic energy of the ejected grains is found proportional to the impact energy as: 
\begin{equation}
E_{ej}\simeq 0.031 E_c \mathrm{ . } \label{eq_enejec} 
\end{equation} 
This unexpected result allows for the definition of an effective restitution coefficient $\rho=\sqrt{E_{ej}/E_c} \approx 0.18$. The small value of $\rho$ confirms the well known and well used fact that granular beds are very efficient dissipating systems.

\section{Discussion} \label{sec_disc} %/ Grain ejection versus crater excavation

In this paper, we have shown that the number of ejected grains $N_{ej}$ obeys a scaling law with the impact energy $E_c$~(see Eq.~(\ref{eq_nejec})). In the literature, the impact energy has been shown to be also the relevant parameter that governs the crater size, in terms of its typical radius and depth~\cite{debruyn03, durian03, zheng, debruyn07}. Let us now relate  these different scaling laws.  
The scaling laws known from Ref.~\cite{debruyn07} for the typical radius $R_{crat}$ and depth $H_{crat}$ of craters are $R_{crat} \propto E_c^{0.23}$  and  $ H_{crat} \propto E_c^{0.21} $. The volume $V$ of craters can thus be also related to the impact energy as $V\sim H_{crat} R^2_{crat}\propto E_c^{0.67}$, with a numerical prefactor that depends on the precise shape of the crater and on the grain and projectile properties.  
From the crater profile measurements of Ref.~\cite{debruyn07} for glass beads, we deduce the precise scaling law for the crater volume: 
\begin{equation}
 V \simeq 3.6\,10^{4} E_c^{0.67} \mathrm{ . } \label{eq_cratervolscal}
\end{equation} 
From this expression for the  crater volume, we can estimate the number of grains that have been ejected as $N_{crat}=V\Phi/v$ where $v=4\pi r^3/3$ is the volume of one grain and $\Phi$ is the solid volume fraction of the granular packing ($\Phi\simeq0.6$). This number of excavated grains follows thus a power law with the impact energy with the exponent $0.67$, that is very close to the exponent $0.70$ measured here for the ejected grains in Eq.~(\ref{eq_nejec}). 
This definitely confirms that the geometrical measurements made on the corona give relevant and quantitative information on the grain ejection dynamics. Writing that the number of ejected grains $N_{ej}$ measured here is equal to the number of excavated grains $N_{crat}$ from~\cite{debruyn07} allows  to bypass the unknow prefactor $w \overline{n}$ in Eq.~(\ref{eq_nej}) and to accurately calculate $N_{ej}$ and $E_{ej}$ shown in Fig.~\ref{FigEnergy}: Imposing the equality $N_{ej} = N_{crat}$  gives  $w \overline{n}\simeq12~mm^{-2}$. 

The scaling laws $u_0\propto E_c^{0.18}$ and $N_{ej}\propto E_c^{0.70}$ obtained here for a large projectile over grain size ratio ($R/r\sim25$-$50$) are close to the results obtained for a small size ratio ($R/r=1$) in~\cite{beladjine}:  $\overline{u_0}\propto E_c^{0.12}$ and $\overline{N_{ej}}\propto E_c^{0.75}$, where $\overline{u_0}$ and $\overline{N_{ej}}$ refer to ejection velocity and number of ejected grains averaged  over many experiments. This suggests that the same scaling laws hold whatever the ratio $R/r$. 
The constant value of the ejection angle equal to about $55^{\circ}\pm5^{\circ}$, is in agreement with previous values measured between $40^{\circ}$ and $60^{\circ}$ in the case of high speed impacts~\cite{cintala,wada}. 

Let us recall that the energy of the ejected grains is shown to correspond to $3\%$ of the impact energy  (Eq.~(\ref{eq_enejec})), that is much larger than the tiny fraction ($0.3\%$ according to~\cite{debruyn07}) required to excavate the crater, i.e. for the grains to move just above the granular surface with zero velocity. 
This is illustrated by the larger values of the maximal height of the corona $H_{max}$ (between $10~mm$ and $40~mm$ in Fig.~\ref{Figeo}a) than the values of crater depth (between $4~mm$ and $10~mm$ in~\cite{debruyn03,debruyn07}) for the same  material properties and impact conditions. 
This suggests to take into account the kinetic energy of the ejected grains  when considering balance of energies. 
However the essential part ($\sim 97\%$) of  the impact energy is dissipated in the granular target through frictional contacts and inelastic collisions.

\section{Conclusion } \label{sec_concl}

The dynamics of grain ejection due to a large sphere impacting a granular material has been  experimentally  investigated through the time  evolution of the corona, mainly in terms of its  geometry. 
Whereas the dimensions  of the corona -- its height, top and bottom diameters -- change with time, the angle formed by its edge with the horizontal granular surface remains constant during the ejection process. 
All these geometrical properties are well described by a simple ballistic model, supporting that grains are quasi-instantaneously ejected at time of impact from the same position in one direction. 
This direction appears to be constant when varying the experimental parameters and equal to about $55^{\circ}$. One may wonder how this angle  changes  with the shape of the impacting projectile. 
By contrast, the typical ejection velocities and number of ejected grains change when varying experimental parameters and are controlled by the impact energy through power laws. 
 The  evaluation of the energy of the ejected grains allows finally  for the calculation of an effective coefficient of restitution characterizing the complex collision process between the impacting sphere and the fine granular target. An important result is that this effective restitution coefficient is constant when varying the experimental parameters and equal to about $0.2$.  

%\acknowledgments{We are grateful to G. Chauvin for his technical help. }

%


\begin{thebibliography}{99}

\bibitem{melosh} H.J.~Melosh, Impact cratering: A geologic process (Oxford University Press, New York, 1989).

\bibitem{cook} M.A.~Cook and K.S.~Mortensen, Impact cratering in granular materials, J. Appl. Phys. {\bf 38}, 13 (1967). 

\bibitem{yamamoto} S.~Yamamoto, K.~Wada, N.~Okabe and T.~Matsui, Transient crater growth in granular targets: An experimental study of low velocity impacts into glass sphere targets, Icarus {\bf 183}, 215-224 (2006). 

\bibitem{debruyn03} A.M.~Walsh, K.E.~Holloway, P.~Habdas and J.R.~de Bruyn, Morphology and scaling of impact craters in granular media, Phys. Rev. Lett. {\bf 91}, 104301 (2003).

\bibitem{durian03}J.S.~Uehara, M.A.~Ambroso, R.P.~Ojha and D.J.~Durian, Low-speed impact craters in loose granular media, Phys. Rev. Lett. {\bf 90}, 194301 (2003).

\bibitem{zheng} X.J.~Zheng, Z.T.~Wang and Z.G.~Qiu, Impact craters in loose granular media, Eur. Phys. J. E {\bf 13}, 321-324  (2004). 

\bibitem{debruyn07} S.J.~de Vet and J.R.~de Bruyn, Shape of impact craters in granular media, Phys. Rev. E {\bf 76}, 041306 (2007). 

\bibitem{debruyn04} J.R.~de Bruyn and A.M.~Walsh, Penetration of spheres into loose granular media, Can. J. Phys. {\bf 82}, 439-446 (2004). 

\bibitem{ciamarra} M.P.~Ciamarra, A.H.~Lara, A.T.~Lee, D.I.~Goldman, I.~Vishik and H.L.~Swinney, Dynamics of drag and force distributions for projectile impact in a granular medium, Phys. Rev. Lett. {\bf 92}, 194301 (2004).

\bibitem{durian05}  M.A.~Ambroso, C.R.~Santore, A.R.~Abate and D.J.~Durian, Penetration depth for shallow impact cratering, Phys. Rev. E {\bf 71}, 051305 (2005). 

\bibitem{hou} M.~Hou, Z.~Peng, R.~Liu, K.~Lu and C.K.~Chan, Dynamics of a projectile penetrating in granular systems, Phys. Rev. E {\bf 72}, 062301 (2005).

\bibitem{durian07} H.~Katsuragi and D.J.~Durian, Unified force law for granular impact cratering, Nature Physics {\bf 3}, 420-423 (2007).

\bibitem{seguin} A.~Seguin, Y.~Bertho and P.~Gondret, Influence of confinement on granular penetration by impact, Phys. Rev. E {\bf 78}, 010301 (2008). 

\bibitem{goldman} D.I.~Goldman and P.~Umbanhowar, Scaling and dynamics of sphere and disk impact into granular media, Phys. Rev. E {\bf 77}, 021308  (2008). 

\bibitem{GDRMidi} G.D.R.~Midi, On dense granular flows, Eur. Phys. J. E {\bf 14}, 341-365 (2004).

\bibitem{mills08} P. Mills, P. Rognon and F. Chevoir, Rheology and structure of granular materials near the jamming transition, Europhys. Lett. {\bf 11}, 619 (2008). 

\bibitem{thoroddsen} S.T.~Thoroddsen and A.Q.~Shen, Granular jets, Phys. Fluids {\bf 13}, 4-6 (2001).

\bibitem{lohse} R.~Mikkelsen, M.~Versluis, E.~Koene, G.-W.~Bruggert, D.~van der Meer, K.~van der Weele, and D.~Lohse,  Granular Eruptions: Void Collapse and Jet Formation, Phys. Fluids  {\bf 14}, S14 (2002).  

\bibitem{royer} J.R.~Royer, E.I.~Corwin, P.J.~Eng and H.M.~Jaeger, Gas-mediated impact dynamics in fine-grained granular materials, Phys. Rev. Lett. {\bf 99}, 038003 (2007). 

\bibitem{caballero} G.A.~Caballero, R.P.~Bergmann, D.~van der Meer, A.~Prosperetti and  D.~Lohse, Role of Air in Granular Jet Formation, 
Phys. Rev. Lett. {\bf 99}, 018001 (2007).

\bibitem{marston} J.O.~Marston, J.P.K.~Seville, Y-V.~Cheun, A.~Ingram, S.P.~Decent, and M.J.H.~Simmon, Effect of packing fraction on granular jetting from solid sphere entry into aerated and fluidized beds, Phys. Fluids {\bf 20}, 023301 (2008).

\bibitem{worthington} A.M.~Worthington, A study of splashes, (Longmans, Green, and Co., London, 1908).

\bibitem{fedorchenko} A.I.~Fedorchenko and A.-B.~Wang, The formation and dynamics of a blob on free and wall sheets induced by a drop impact on surfaces, Phys. Fluids {\bf 16}, 3911-3920 (2004). 

\bibitem{yarin} A.L.~Yarin, Drop impact dynamics: Splashing, spreading, receding, bouncing, Ann. Rev. Fluid Mech. {\bf 38}, 159-192 (2006).

\bibitem{ogale} S.B.~Ogale, S.R.~Shinde, P.A.~Karve, A.S.~Ogale, A.~Kulkarni, A.~Athawale, A.~Phadke and R.~Thakurdas, Impact-induced splash and spill in a quasi-confined granular medium, Physica A {\bf 363} 187-197 (2006). 

\bibitem{boudet} J-F.~Boudet, Y.~Amarouchene and H.~Kellay, Dynamics of impact cratering in shallow sand layers, Phys. Rev. Lett. {\bf 96}, 158001 (2006). 

\bibitem{cintala} M.J.~Cintala, L.~Berthoud, and F.~Horz, Ejection-velocity distribution from impact into coarse-grained sand, Meteor. Planet. Sci. {\bf 34}, 605-623 (1999). 

\bibitem{beladjine} D.~Beladjine, M.~Ammi, L.~Oger and A.~Valance, Collision process between an incident bead and a three-dimensional granular packing, Phys. Rev. E {\bf 75}, 061305 (2007).  

\bibitem{tsimring} L.S.~Tsimring and D.~Volfson, Modeling of impact cratering in granular media, in Powders and Grains (A. A. Balkema Publishers, Stuttgart, 2005), pp 1215-1218.

\bibitem{wada} K.~Wada, H.~Senshu and T.~Matsui, Numerical simulation of  impact cratering on granular material, Icarus {\bf 180}, 528-545 (2006). 

\bibitem{oger} L.~Oger, M.~Ammi,  A.~Valance and D.~Beladjine, Discrete element method studies of the collision of one rapid sphere on 2D and 3D packings, Eur. Phys. J. {\bf 17}, 467-476 (2005).

\bibitem{crassous} J.~Crassous, D.~Beladjine and A.~Valance , Impact of a projectile on a granular medium described by a collision model, Phys. Rev. Lett. {\bf 99}, 248001 (2007).

\bibitem{bourrier} F.~Bourrier, F.~Nicot and F.~Darve, Physical processes within a 2D granuler layer during an impact, Granular Matter {\bf 10}, 6 (2008).

\end{thebibliography}
\end{document}